\documentclass[a4paper,11pt]{article}
\pdfoutput=1 

\usepackage{jheppub} 

\usepackage[T1]{fontenc} 
\usepackage{feynmp-auto}

\title{\boldmath $O(N)$ model in Euclidean de Sitter space: beyond the leading infrared approximation}

\author[a]{Diana  L\'opez Nacir}
\author[b]{Francisco D. Mazzitelli}
\author[b]{Leonardo G. Trombetta}

\affiliation[a]{Theoretical Physics Department, CERN,\\CH-1211 Gen\`eve 23, Switzerland.}
\affiliation[b]{Centro At\'omico Bariloche, Instituto Balseiro and CONICET,\\ Comisi\'on Nacional de Energ\'\i a At\'omica, Av. Bustillo 9500, R8402AGP Bariloche, Argentina.}

\emailAdd{diana.laura.lopez.nacir@cern.ch}
\emailAdd{fdmazzi@cab.cnea.gov.ar}
\emailAdd{lgtrombetta@cab.cnea.gov.ar}

\abstract{We consider an $O(N)$ scalar field model with quartic interaction in $d$-dimensional Euclidean de Sitter space. In order to avoid the problems of the standard perturbative calculations for light and massless fields,  we generalize to the $O(N)$ theory a systematic method introduced previously for a single field, which treats the zero modes exactly and the nonzero modes perturbatively. We compute the two-point functions taking into account not only the leading infrared  contribution, coming from the self-interaction of the zero modes, but also corrections  due to the interaction of the ultraviolet modes. For the model defined in the corresponding Lorentzian de Sitter spacetime, we obtain the two-point functions by analytical continuation. We point out that a partial resummation of the leading secular terms (which necessarily involves nonzero modes) is required
 to obtain a decay at large distances for massless fields.
 We implement this resummation along with a systematic double expansion in an effective coupling constant $\sqrt\lambda$ and in $1/N$. 
 We explicitly perform the calculation up to the next-to-next-to-leading order in $\sqrt\lambda$ and up to next-to-leading order in $1/N$.  The results reduce to those known in the leading infrared approximation. We also show that they coincide with the ones obtained directly in Lorentzian de Sitter spacetime in the large $N$ limit, provided the same renormalization scheme is used.}
 
\begin{document} 

\hspace{4.3in} \mbox{CERN-TH-2016-126}

\maketitle
\flushbottom


\section{Introduction}

The study of interacting quantum fields in de Sitter geometry is of interest for a variety of reasons.  In inflationary models, interactions  could lead to non-Gaussianities
in the cosmic microwave background. Quantum effects could also contribute to the dark energy, and explain, at least partially, the present acceleration of the universe.
From a conceptual point of view, being maximally symmetric, de Sitter geometry allows for a number of explicit analytic calculations that illustrate the role
of the curved background on the dynamics of the quantum field, and also the backreaction of the quantum fields on the geometry.

Although at first sight quantum field theory (QFT) in de Sitter geometry should be simpler than on other cosmological backgrounds, the exponential expansion of the metric
produces an effective growth in the couplings. Indeed, when considering a $\lambda\phi^4$ theory, a diagram with $L$ loops will be proportional to 
$(\lambda H^2/m^2)^L$, where $m$ is the mass of the quantum field and $H$ the Hubble constant \cite{Burgess}. Therefore, the usual perturbative calculations break down for
light ($m^2\ll H^2$) quantum fields. In the  massless limit, minimally coupled scalar fields do not admit  a de Sitter invariant vacuum state due to infrared
(IR) divergences, and the two-point functions do not respect the symmetries of the classical theory \cite{Allen}.   
 
 There have been several attempts to cure the IR problem, all of them introducing some sort of nonperturbative approach.
The well known stochastic approach \cite{Starobinsky} assumes that the long distance fluctuations of the quantum field behave as 
a random classical field,  that satisfies a Langevin equation. From the corresponding Fokker-Planck equation one can obtain finite and de  Sitter invariant 
correlation functions  even for  minimally coupled massless fields and fields with negative mass-squared, provided there are interaction terms that stabilize the potential. 
The formalism has  proven very useful in understanding the IR effects and, in particular, the spontaneous symmetry breaking phenomena (see for instance \cite{Starobinsky,Prokopec}).  However,
it does not provide a systematic way of accounting for the interactions of the short distance fluctuations, which are generically neglected, nor a justification for the 
  classical treatment of the field beyond the leading IR approximation. 

  In the context of QFT  in Lorentzian de Sitter spacetime,
 many of the nonperturbative approaches are based in mean field (Hartree) \cite{Hartree1, Hartree2} or large $N$ approximations \cite{LargeN}, which in turn can be formulated through
 the 2PI effective action (2PIEA) \cite{CJT}.
Other  approaches consider the analysis of the Schwinger-Dyson 
 equations (or their nonequilibrium counterpart, the Kadanoff-Baym equations), combined with a separation of long and short 
 wavelengths using a physical momentum decomposition \cite{Gautier}. There are also calculations based on the dynamical renormalization group,
 and on the exact renormalization group equation for the effective potential \cite{Guilleux}.   In relation to the  development of a method for computing correlation functions in a  systematically improvable way, the difficulties one finds with these approaches are mainly of technical nature.
 
 An alternative and simpler approach emerges for quantum fields in Euclidean de Sitter space. As this is a compact space (a sphere of radius
 $H^{-1}$) the modes of a quantum field are discrete, and the origin of the IR problems can be traced back to the zero mode \cite{Rajaraman1}. 
 This important observation  suggests itself the solution: the zero mode should be treated exactly while the nonzero modes (UV modes in what follows) could be treated
 perturbatively.  Using this idea, it has been shown that in the massless $\lambda\phi^4$ theory the interaction turns on a dynamical squared mass for the field, that in the 
 leading IR limit is proportional to  $\sqrt{\lambda} H^2$.  Indeed, in the Euclidean space, this mass cures the IR problems. 
   Moreover, a systematic perturbative procedure for calculating  the $n$-point functions has been delineated    in Ref.~\cite{BenekeMoch}, where it was shown that, for massless fields, the effective coupling is $\sqrt\lambda$ instead of $\lambda$. It has been pointed out  that this procedure together with an analytical continuation  could  be used  to  
    cure  some IR problems also in the Lorentzian de Sitter spacetime, and  in particular to obtain $n$-point functions that respect the de Sitter symmetries.
 However,  so far explicit calculations have been restricted to obtaining corrections to the variance of the zero mode, which has no analog quantity in the Lorentzian de Sitter spacetime. 
 In particular, an explicit calculation of the inhomogeneous two-point functions 
of the scalar field that can lead to 
  two-point functions  respecting  the de Sitter symmetries after  analytical continuation  is still lacking. To perform that explicit calculation is one of the main goals of this paper.

 The relation between the different approaches (Lorentzian, Euclidean, stochastic), has been the subject of several works. 
 For example, recently, it has been shown that the stochastic and the {\it in-in} Lorentzian calculations are equivalent at the level of Feynman rules for massive fields, when computing equal-time correlation functions in the leading IR approximation \cite{Garbrecht},  in agreement with previous arguments \cite{Tsamis}.   The evaluation of the field variance in the large $N$ limit  gives the same result in the three approaches, up to next to leading order (NLO) in $1/N$, in the IR limit \cite{Gautier}.  The connection between the Feynman diagrams for computing correlation functions in the {\it in-in}  formalism and  the analytically continued ones obtained in the Euclidean space has been described in detail in \cite{Marolf,Hollands} for massive fields.   Similar arguments have been used in \cite{HollandsMassless} to generalize the connection to  the  massless case when the zero mode is treated nonperturbatively as proposed in \cite{Rajaraman1}. In the latter case,  the inhomogeneous IR behavior of the the correlation functions is still unclear (they could grow at large distances \cite{HollandsMassless}), and  to understand this it is necessary to go beyond the leading IR approximation. 
   It is worth to highlight that beyond the leading IR approximation a comparison of the results obtained by different approaches  necessary requires the use of equivalent regularization methods and renormalization schemes.

 In this paper we will pursue the approach of Ref.~\cite{BenekeMoch}, providing a generalization to the case of $O(N)$ scalar field theory, and including
 a discussion of the renormalization process. We will present a detailed calculation of the corrections to the two-point functions up to second order in the 
 parameter $\sqrt \lambda$. As we will see, the zero mode part of the two-point functions, that also receives UV corrections, determines the quadratic part of the effective potential. Being positive for all values of $d$ and $N$, it implies that spontaneous symmetry breaking does not occur in the $O(N)$ models for $\lambda\ll 1$. We will  check that, in the leading large $N$ limit, the Euclidean results are fully consistent with the Lorentzian ones, even
 beyond the leading IR approximation. We will also show that  the corrections of order $1/N$ of our result coincide  in the leading IR approximation with the ones  of Ref.~\cite{Gautier},  which are  the most precise results known so far for this model  and  were obtained working   in the leading IR approximation and directly in the framework of the QFT in Lorentzian spacetime. Our results improve on those by including systematically
 the corrections coming from the interactions of both IR and UV sectors. 
  
 The paper is organized as follows. In Section II we present our model and describe the systematic perturbative calculation of Ref.~\cite{BenekeMoch}. 
 We compute the homogeneous part of the two-point functions, related to the quadratic part of the effective potential, in the leading IR approximation (i.e. neglecting the contributions coming from the UV modes). In Section III we evaluate the full two-point functions of the theory, up to corrections of second order from the interaction with the UV modes, including the counterterms needed for renormalization. We find out that, although this framework deals with the IR divergences by giving a dynamical mass to the zero modes, the behavior of the two-point functions at large distances  still has IR problems for massless fields. In Section IV we analyze in detail the massless case. We perform a large $N$ expansion of the results of  Section III and  compare the Euclidean and Lorentzian results. Section V is concerned with an extension of the nonperturbative treatment to perform a resummation
of the leading IR secular terms, in order to recover the proper decay of the two-point functions at large distances. This resummation is combined with a systematic expansion both in $\sqrt\lambda$ and $1/N$.
Finally, in Section VI we describe the main conclusions of our work. Several  Appendices contain some details of the calculations.

\section{Euclidean de Sitter Space}

In this section we describe the methods of QFT in the Euclidean de Sitter
 space developed in \cite{Rajaraman1} and \cite{BenekeMoch} and generalize them for the case of a $O(N)$-symmetric model with Euclidean action
\begin{equation}
 S = \int d^d x \sqrt{g} \left[ \frac{1}{2} \phi_a \left( -\square + m^2 \right) \phi_a + \frac{\lambda}{8N} ( \phi_a \phi_a )^2 \right],
\end{equation}
where $\square = \frac{1}{\sqrt{g}} \partial_\mu \left( \sqrt g g^{\mu \nu} \partial_\nu \right)$,  $\phi_a$ denotes the components of an element of the adjoint representation of the $O(N)$ group, with $a=1,..,N$, and the sum over repeated indices is implied. Also, a possible nonminimal coupling with the curvature $\xi$ is included in the mass parameter $m^2 = \tilde{m}^2 + \xi d(d-1) H^2$. In $d$ dimensions the field coupling constant has units of $H^{4-d}$, thus can be expressed as $\lambda = \mu^{4-d} \lambda_4 $, with $\lambda_4$ a dimensionless constant and $\mu$ a scale with mass dimensions.

Euclidean de Sitter
 space is obtained from Lorentzian de Sitter
 space in global coordinates by performing an analytical continuation $t \to -i ( \tau - \pi/2H )$ and a compactification in imaginary time $\tau = \tau + 2\pi H^{-1}$. The resulting metric is that of a $d$-sphere of radius $H^{-1}$
\begin{equation}
 ds^2 = H^{-2} \left[ d \theta^2 + \sin(\theta)^2 d\Omega^2 \right],
\end{equation}
where $\theta = H\tau$. Due to the symmetries and compactness of this space, the field can be expanded in $d$-dimensional spherical harmonics
\begin{equation}
 \phi_a(x) = \sum_{\vec{L}} \phi_{\vec{L},a} Y_{\vec{L}}(x),  
\end{equation}
and then the free scalar propagator of mass $m$ is, in the symmetric phase, 
\begin{eqnarray}
 G^{(m)}_{ab}(x,x') &=& \delta_{ab} G^{(m)}(x,x') = \delta_{ab} H^d \sum_{\vec{L}} \frac{Y_{\vec{L}}(x) Y^*_{\vec{L}}(x')}{H^2 L(L+d-1) + m^2},
\end{eqnarray}
where the superscript indicates the mass. The $\vec{L}=\vec{0}$ contribution $G_0^{(m)} = |Y_{\vec{0}}|^2 H^d/m^2$ is clearly responsible for the infrared divergence in the correlation functions of the scalar field for $m^2 \to 0$. We split $\phi_a(x) =  \phi_{0a} + \hat{\phi}_a(x)$ in order to treat the constant zero modes $\phi_{0a}$ separately from the inhomogeneous parts $\hat{\phi}_a(x)$. This prompts to separate the propagator as well,
\begin{equation}
 G^{(m)}(x,x') = G_0^{(m)} + \hat{G}^{(m)}(x,x'),
\end{equation}
where now $\hat{G}^{(m)}$ has the property of being finite in the infrared ($m^2 \to 0$).  

The interaction part of the action takes the following form:
\begin{equation}
 S_{int} = \frac{\lambda V_d}{8N} |\phi_0|^4 + \tilde{S}_{int}[\phi_{0a},\hat{\phi}_a].
\end{equation}
Here $\vert\phi_0\vert^2=\phi_{0a} \phi_{0a}$ and $V_d$ is the total volume of Euclidean de Sitter
 space in $d$-dimensions, which thanks to the  compactification is finite and equal to the hypersurface area of a $d$-sphere
\begin{equation}
 V_d = \int d^d x \sqrt{g} = \frac{2 \pi^{\frac{d+1}{2}}}{\Gamma\left( \frac{d+1}{2} \right) H^d} = \frac{1}{|Y_{\vec{0}}|^2 H^d},
\end{equation}
where  $\Gamma$ is Euler's Gamma function. The explicit form of $ \tilde{S}_{int}$ will be written below.

In order to compute the quantum correlation functions of the theory we define the generating functional in the presence of sources $J_{0a}$ and $\hat{J}_a$,
\begin{eqnarray}
   Z[J_0, \hat{J}] &=& \mathcal{N} \int d^N \phi_0 \int \mathcal{D}\hat{\phi} \, e^{-S - \int_x  (J_{0a} \phi_{0a} + \hat{J}_a \hat{\phi}_a )} \notag \\
   &=& exp\left( -\tilde{S}_{int}\left[\frac{\delta}{\delta J_0},\frac{\delta}{\delta \hat{J}}\right] \right) Z_0[J_0] \hat{Z}_{f}[\hat{J}],
   \label{Z-int}
\end{eqnarray}
where we introduced the shorthand notation $\int_x = \int d^d x \sqrt{g}$. Also, in the second line $Z_0[J_0]$ is defined as the generating functional of the theory with the zero modes alone. This part gives the leading infrared contribution and can be exactly computed in several interesting cases following Ref.~\cite{Rajaraman1}. Note that, as the zero modes are constant on the sphere,  their kinetic terms vanish, and $Z_0[J_0]$ involves only ordinary integrals.

The effective potential gives valuable information about how the quantum fluctuations around a background field $\bar{\phi}$ influence its behavior. We are interested in particular in the generation of a dynamical mass due to quantum effects. Up to quadratic order it can be shown that (see Appendix \ref{app-eff-pot}),
\begin{eqnarray}
 V_{eff} (\bar{\phi}_0) &=& V_0 + \frac{1}{2} \frac{N}{V_d \langle \phi_0^2 \rangle} |\bar{\phi}_0|^2 + \mathcal{O}(|\bar{\phi}_0|^4). \label{eff-pot-quad}
\end{eqnarray}
This is an exact property of the Euclidean theory valid for all $N$ and $\lambda$, which shows that the dynamical mass is  related to the inverse of the variance of the zero modes as \begin{equation}
 m_{dyn}^2 =\frac{N}{V_d \langle \phi_0^2 \rangle}. \label{full-mdyn-from-Veff}
\end{equation}

At the leading infrared order the interaction between the infrared and ultraviolet modes in Eq.~\eqref{Z-int} can be neglected, and
\begin{equation}
 \langle \phi_0^2 \rangle_0 = \delta_{ab} \frac{\delta^2 Z_0[J_0]}{\delta J_{0a} \delta J_{0b}} \Bigg|_{J_0=0} = \frac{\int d^N\phi_0 \, \phi_0^{2} e^{-V_d \left[\frac{\lambda}{8N} \phi_0^4 + \frac{m^2}{2} \phi_0^2 \right]}}{\int d^N\phi_0 \, e^{-V_d \left[ \frac{\lambda}{8N} \phi_0^4 + \frac{m^2}{2} \phi_0^2 \right]}}. 
\end{equation}
For a vanishing tree level mass $m=0$, the integrals on the right-hand side can be computed exactly leading to a dynamical mass,
\begin{equation}
 m_{dyn,0}^2 = \sqrt{\frac{N \lambda}{2V_d}} \frac{1}{2} \frac{\Gamma\left[ \frac{N}{4} \right]}{\Gamma\left[ \frac{N+2}{4} \right]}. \label{mdyn0-allN}
\end{equation}
For $N = 1$, we recover the result of \cite{Rajaraman1}, which is also the one from the stochastic approach \cite{Starobinsky},
\begin{equation}
 m_{dyn,0}^2 \Bigg|_{N=1} = \frac{\sqrt{3\lambda_4} H^2}{8\pi} \frac{\Gamma\left[ \frac{1}{4} \right]}{\Gamma\left[ \frac{3}{4} \right]},
\end{equation}
where we evaluated in $d=4$, $V_4 = 8\pi^2/3H^4$.

In the following section, we will compute the two-point functions 
of the full scalar field including up  to the second perturbative correction coming from the UV modes. 

\section{Corrections from the UV modes to the two-point functions}

Corrections to the leading order result come from expanding the exponential with $\tilde{S}_{int}$ in Eq.~\eqref{Z-int}. The explicit expression for the interaction part of the action that involves the UV modes is 
\begin{eqnarray}
 \tilde{S}_{int} &=& \frac{\lambda}{8N} \int d^d x \sqrt{g} \, \Biggl[ 2 A_{abcd} \phi_{0a} \phi_{0b} \hat{\phi}_c \hat{\phi}_d + 4 \delta_{ab} \delta_{cd} \phi_{0a} \hat{\phi}_{b} \hat{\phi}_c \hat{\phi}_d + \delta_{ab} \delta_{cd} \hat{\phi}_a \hat{\phi}_b \hat{\phi}_c \hat{\phi}_d \Biggr],
 \label{S-int-tilde}
\end{eqnarray}
where $A_{abcd}$ is the totally symmetric 4-rank tensor
\begin{equation}
A_{abcd} = \delta_{ab} \delta_{cd} + \delta_{ac} \delta_{bd} + \delta_{ad} \delta_{bc}.
\end{equation}
The terms linear in $\hat{\phi}$ that would appear when splitting both the mass and the interaction terms vanish, since $\int d^d x \sqrt{g} Y_{\vec{L}}(x) = 0$ for $L>0$.

As a guiding principle for computing the perturbative corrections in this section, we recall that for a massless minimally coupled field, $\langle\phi_0^{2p} \rangle_0 \sim \lambda^{-p/2}$ (see Eqs.~\eqref{full-mdyn-from-Veff} and \eqref{mdyn0-allN}, and Eq.~\eqref{phi_0-2p-massless} below), and therefore we will have a perturbative expansion in powers of $\sqrt{\lambda}$ (in contrast, the correlation functions of the zero modes start at $\lambda^0$ when $m \neq 0$, and therefore the order at which each perturbative term contributes changes with respect to the massless case). With this in mind, here we will keep terms that will be at most of order $\lambda$ when $m=0$.

The first correction to the generating functional comes from expanding the exponential linearly and keeping the first term of Eq.~\eqref{S-int-tilde}. Following the standard  procedure,
the generating functional at NLO reads
\begin{eqnarray}
 Z[J_0,\hat{J}] &=& Z_0[J_0] \hat{Z}_{f}[\hat{J}] - \frac{\lambda}{4N} A_{abcd} \frac{ \delta^2 Z_0 [J_0] }{\delta J_{0a} \delta J_{0b}} \int d^d x \sqrt{g} \frac{ \delta^2 \hat{Z}_{f}[\hat{J}] }{\delta \hat{J}_c(x) \delta \hat{J}_d(x)}. 
\label{Z-int-NLO}
\end{eqnarray}

The next to NLO (NNLO) correction  has two contributions. The first one is given by the square of the interaction term considered before, as it comes from expanding the exponential up to quadratic order. The second one is the last term in Eq.~\eqref{S-int-tilde} at linear order. These NNLO contributions to the generating functional are given by
\begin{eqnarray}
 &+& \frac{1}{2} \frac{\lambda^2}{16N^2} A_{abcd} A_{efgh} \frac{ \delta^4 Z_0 [J_0] }{\delta J_{0a} \delta J_{0b} \delta J_{0e} \delta J_{0f}} \int d^d x \sqrt{g} \int d^d x' \sqrt{g'} \frac{ \delta^4 \hat{Z}_{f}[\hat{J}] }{\delta \hat{J}^4_{cdgh}(x,x,x',x')}, \label{Z-int-NNLO1}
\end{eqnarray}
and
\begin{equation}
 - \frac{\lambda}{8N} Z_0[J_0] \delta_{ab} \delta_{cd} \int d^d x \sqrt{g} \frac{ \delta^4 \hat{Z}_{f}[\hat{J}] }{\delta \hat{J}^4_{abcd}(x,x,x,x)},
\label{Z-int-NNLO2}
\end{equation}
respectively. Here, and in what follows, we use the notation 
$$\delta\hat{J}^k_{a_1...a_k}(x_1,...,x_k) = \delta\hat{J}_{a_1}(x_1)...\delta\hat{J}_{a_k}(x_k)$$ 
as a shorthand in  the functional derivatives.


We split the  two-point functions of the total fields $\phi_a$ into UV and IR parts 
\begin{equation}
 \langle \phi_{a}(x) \phi_{b}(x') \rangle = \langle \phi_{0a} \phi_{0b} \rangle + \langle \hat{\phi}_{a}(x) \hat{\phi}_{b}(x') \rangle\, ,
\end{equation}
where the cross-terms vanish by orthogonality.
In what follows we compute each part separately. 


\subsection{UV part of the two-point functions}

Starting with the UV part, we calculate the two point functions of $\hat{\phi}_a$ by taking two functional derivatives of $Z[J_0,\hat{J}]$ with respect to $\hat{J}_a(x)$,
\begin{eqnarray}
 \langle \hat{\phi}_{a}(x) \hat{\phi}_{b}(x') \rangle &=& \frac{1}{Z[0,0]} \frac{ \delta^2 Z[J_0,\hat{J}] }{\delta \hat{J}_a(x) \delta \hat{J}_b(x')} \Bigg|_{J_0,\hat{J}=0}
  \label{2-pt-UV}
\end{eqnarray}
where the factor $Z[0,0]^{-1}$ takes care of the normalization of the interacting theory: 
\begin{eqnarray}
Z[0,0]^{-1} &=& 1 + \frac{\lambda}{4N} (N+2) \langle \phi_0^2 \rangle_0 V_d [\hat{G}^{(m)}] + \frac{\lambda}{8} (N+2) V_d [\hat{G}^{(m)}]^2 \notag \\
&&+\frac{\lambda^2}{16N^2} (N+2)^2 \langle \phi_0^2 \rangle_0^2 V_d^2 [\hat{G}^{(m)}]^2 \notag \\
&&- \frac{\lambda^2 \langle \phi_0^4 \rangle_0}{32N^2} \Biggl[ (N+2)^2 V_d^2 [\hat{G}^{(m)}]^2 +2(N+8) \iint_{x,x'} \hat{G}^{(m)}(x,x')^2 \Biggr]\, .
\label{Z00}
\end{eqnarray}
Here $[\hat{G}^{(m)}]$ denotes the coincidence limit of the free UV propagator with mass $m$, which is independent of $x$ by de Sitter invariance, and we used $A_{abcd} \delta_{cd} = (N+2) \delta_{ab}$. Furthermore, to arrive at this expression we have used Eqs.~\eqref{Zf2} and \eqref{Zf4} to write the derivatives of $\hat{Z}_{f}[\hat{J}]$ in terms of free propagators, relying on the fact that it is a free generating functional. On the other hand, for the derivatives of $Z_0[J_0]$, we have
\begin{eqnarray}
\frac{\delta^2 Z_0 [J_0] }{\delta J_{0a} \delta J_{0b}} \Bigg|_{J_0=0} &=& \langle \phi_{0a} \phi_{0b} \rangle_0 = \delta_{ab} \frac{\langle \phi_0^2 \rangle_0}{N}, \label{2-pt-zero} \\
\frac{\delta^4 Z_0 [J_0] }{\delta J_{0a} \delta J_{0b} \delta J_{0c} \delta J_{0d}} \Bigg|_{J_0=0} &=& A_{abcd} \frac{\langle \phi_0^4 \rangle_0}{N(N+2)}, \label{4-pt-zero}
\end{eqnarray}
with $\langle \phi_{0a} \phi_{0b} \rangle_0$ the exact two-point functions 
of the zero modes in the absence of the UV modes. By exact we mean that we include the self-interaction nonperturbatively. For convenience, in the last equality of Eq.~\eqref{2-pt-zero} we expressed the result in terms of its trace with respect to the internal $O(N)$ indices. In general, any of the traced exact $n$-point functions of zero modes at leading order can be expressed in terms of ordinary integrals. For even $n=2p$
we have
\begin{equation}
 \langle \phi_0^{2p} \rangle_0 = \frac{\int d^N\phi_0 \, \phi_0^{2p} e^{-V_d \left[\frac{\lambda}{8N} \phi_0^4 + \frac{m^2}{2} \phi_0^2 \right]}}{\int d^N\phi_0 \, e^{-V_d \left[ \frac{\lambda}{8N} \phi_0^4 + \frac{m^2}{2} \phi_0^2 \right]}}, 
\end{equation}
while they vanish for odd $n$.

To evaluate the second derivative of $Z[J_0,\hat{J}]$,
we can set $J_0=0$ in Eq.~\eqref{2-pt-UV} and compute
\begin{eqnarray}
 \frac{ \delta^2 Z[0,\hat{J}] }{\delta \hat{J}_a(x) \delta \hat{J}_b(x')} \Bigg|_{\hat{J}=0} &=& \frac{ \delta^2 \hat{Z}_{f}[\hat{J}]}{\delta \hat{J}_a(x) \delta \hat{J}_b(x') } \Bigg|_{\hat{J}=0} - \frac{\lambda}{4N} (N+2) \langle \phi_0^2 \rangle_0 \int_{z} \frac{ \delta^4 \hat{Z}_{f}[\hat{J}]}{\delta \hat{J}^4_{abcd}(x,x',z,z) } \Bigg|_{\hat{J}=0} \notag \\
 &&+ \frac{\lambda^2 \langle \phi_0^4 \rangle_0}{32N^3} \left[ (N+4) \delta_{cd} \delta_{ef} + 4\frac{A_{cdef}}{(N+2)} \right] \iint_{y,z} \frac{ \delta^6 \hat{Z}_{f}[\hat{J}] }{\delta \hat{J}^6_{abcdef}(x,x',y,y,z,z)} \Bigg|_{\hat{J}=0} \notag \\
 &&-\frac{\lambda}{8N} \delta_{cd} \delta_{ef} \int_{z} \frac{ \delta^6 \hat{Z}_{f}[\hat{J}] }{\delta \hat{J}^6_{abcdef}(x,x',z,z,z,z)} \Bigg|_{\hat{J}=0}\, . \label{Z-2hatJ}
\end{eqnarray}
Once again we make use of Eqs.~\eqref{Zf2}, \eqref{Zf4} and similar expressions in Eqs.~\eqref{Zf6a} and \eqref{Zf6b} to replace the derivatives of $\hat{Z}_f[\hat{J}]$ in Eq.~\eqref{Z-2hatJ}. Then we multiply by the normalization factor in Eq.~\eqref{Z00}. The result is:   
\begin{eqnarray}
 \langle \hat{\phi}_{a}(x) \hat{\phi}_{b}(x') \rangle &=& \delta_{ab} \Biggl\{ \hat{G}^{(m)}(x,x') + \Biggl[ - \frac{\lambda (N+2)}{2N^2} \langle \phi_0^2 \rangle_0 -\frac{\lambda}{2N}(N+2) [\hat{G}^{(m)}] \notag \\
 &&+\frac{\lambda^2}{8N^3}(N+2)^2 V_d [\hat{G}^{(m)}] \left( \langle \phi_0^4 \rangle_0 - \langle \phi_0^2 \rangle_0^2 \right) \Biggr]  \int_{z} \hat{G}^{(m)}(x,z) \hat{G}^{(m)}(z,x') \notag \\
 &&+ \frac{\lambda^2}{4N^3}(N+8) \langle \phi_0^4 \rangle_0 \int_{y,z} \hat{G}^{(m)}(x,y) \hat{G}^{(m)}(y,z) \hat{G}^{(m)}(z,x') \label{2-pt-UV2} 
 \Biggr\}.\,\,\,\,\,\,\,\,
\end{eqnarray}
 
Eq.~\eqref{2-pt-UV2} needs the inclusion of counterterms to renormalize the divergences present in $[\hat{G}^{(m)}]$. The details of this process are given in the Appendix \ref{app-ren}. Splitting
\begin{equation}
 [\hat{G}^{(m)}] = [\hat{G}^{(m)}]_{div} + [\hat{G}^{(m)}]_{ren},
\end{equation}
it can be seen that Eq.~\eqref{2-pt-UV2} can be made finite by a mass counterterm of the form
\begin{equation}
 \delta m^2 = - \frac{\lambda}{2N} (N+2) [\hat{G}^{(m)}]_{div}. \label{delta-m-euclidean}
\end{equation}

The expression for the UV part of the propagator can be simplified considerably using that the integrals of free UV propagators in Euclidean space in Eq.~\eqref{2-pt-UV2} can be expressed in terms of derivatives of a single propagator with respect to its mass. This is  shown in the Appendix \ref{app-int-UVprop}:
\begin{eqnarray}
 \int_{z} \hat{G}^{(m)}(x,z) \hat{G}^{(m)}(z,x') &=& - \frac{\partial \hat{G}^{(m)}(x,x')}{\partial m^2}, \label{G2-dGdm} \\
 \iint_{y,z} \hat{G}^{(m)}(x,y) \hat{G}^{(m)}(y,z) \hat{G}^{(m)}(z,x') &=& \frac{1}{2} \frac{\partial^2 \hat{G}^{(m)}(x,x')}{\partial (m^2)^2}, \label{G3-d2Gdm2}
\end{eqnarray}
and thus Eq.~\eqref{2-pt-UV2} reads, after renormalization, 
\begin{eqnarray}
 \langle \hat{\phi}_{a}(x) \hat{\phi}_{b}(x') \rangle = \delta_{ab} &\Biggl\{& \hat{G}^{(m)}(x,x')
 + \Biggl[ \frac{\lambda (N+2)}{2N^2} \langle \phi_0^2 \rangle_0 +\frac{\lambda}{2N}(N+2) [\hat{G}^{(m)}]_{ren} \notag \\
 &&-\frac{\lambda^2}{8N^3}(N+2)^2 V_d [\hat{G}^{(m)}]_{ren} \left( \langle \phi_0^4 \rangle_0 - \langle \phi_0^2 \rangle_0^2 \right) \Biggr] \frac{\partial \hat{G}^{(m)}(x,x')}{\partial m^2} \notag \\
 &&+ \frac{\lambda^2}{8N^3}(N+8) \langle \phi_0^4 \rangle_0  \frac{\partial^2 \hat{G}^{(m)}(x,x')}{\partial (m^2)^2} \label{2-pt-UV-ren} 
 \Biggr\}.
\end{eqnarray}
Na\"ively, this result looks like a Taylor expansion of a free UV propagator with respect to the mass squared, around the mass parameter $m^2$. However, as we will see later, while this  is indeed the case in  the large-$N$ limit, it does not hold anymore when the $1/N$ corrections are included.


\subsection{IR part of the two-point functions}

We now calculate the IR part of the two-point functions 
by taking two derivatives of the generating functional $Z [J_0,\hat{J}]$ with respect to $J_{0}$,
\begin{equation}
 \langle \phi_{0a} \phi_{0b} \rangle = \frac{1}{Z[0,0]} \frac{ \delta^2 Z [J_0,\hat{J}] }{\delta J_{0a} \delta J_{0b}} \Bigg|_{J_0,\hat{J}=0}.
\end{equation}
It is useful to first set $\hat{J}=0$ and take the derivatives afterwards: 
\begin{eqnarray}
\frac{ \delta^2 Z [J_0,\hat{J}] }{\delta J_{0a} \delta J_{0b}} \Bigg|_{J_0,\hat{J}=0} &=& \frac{\delta_{ab}}{N} \Biggl\{ \langle \phi_0^2 \rangle_0 \left[ 1 - \frac{\lambda}{8}(N+2) V_d [\hat{G}^{(m)}]^2 \right] - \frac{\lambda}{4N} (N+2) \langle \phi_0^4 \rangle_0 V_d [\hat{G}^{(m)}]  \notag \\
&&+ \frac{\lambda^2}{32N^2} \langle \phi_0^6 \rangle_0 \Biggl[ (N+2)^2 V_d^2 [\hat{G}^{(m)}]^2
+ 2(N+8) \iint_{x,x'} \hat{G}^{(m)}(x,x')^2 \Biggr] \Biggr\}, \notag \\
&&\label{Z-2J0}
\end{eqnarray}
where   we made use of Eqs.~\eqref{2-pt-zero} and \eqref{4-pt-zero}, and that
\begin{equation}
\delta_{cd} \delta_{ef} \frac{\delta^6 Z_0 [J_0] }{\delta J_{0a} \delta J_{0b} \delta J_{0c} \delta J_{0d} \delta J_{0e} \delta J_{0f}} \Bigg|_{J_0=0} = \delta_{ab} \frac{\langle \phi_0^6 \rangle_0}{N}. \label{6-pt-zero}
\end{equation}
In the last term of Eq.~\eqref{Z-2J0} we can make the replacement $\int_{x'} \hat{G}^{(m)}(x,x')^2 = -\partial [\hat{G}^{(m)}]/\partial m^2$ by virtue of Eq.~\eqref{G2-dGdm}. This shows that  this expression has, besides $[\hat{G}^{(m)}]$, another divergent quantity $\partial [\hat{G}^{(m)}]/\partial m^2$. We split the latter  as before
\begin{eqnarray}
 \frac{\partial [\hat{G}^{(m)}]}{\partial m^2} &=& \left( \frac{\partial [\hat{G}^{(m)}]}{\partial m^2}\right)_{div} + \left( \frac{\partial [\hat{G}^{(m)}]}{\partial m^2}\right)_{fin},
\end{eqnarray}
and following the details of the Appendix \ref{app-ren} we obtain the renormalized expression. The renormalization involves a new counterterm to compensate for this divergence, namely
\begin{eqnarray}
 \delta \lambda &=& -\frac{\lambda^2}{2N} (N+8) \left( \frac{\partial [\hat{G}^{(m)}]}{\partial m^2}\right)_{div}. \label{delta-lambda-euclidean}
\end{eqnarray}
%
Finally, we can write down the renormalized two-point functions for the zero modes
\begin{eqnarray}
 \langle \phi_{0a} \phi_{0b} \rangle &=& \delta_{ab} \Biggl\{ \frac{\langle \phi_0^2 \rangle_0}{N} + \frac{\lambda}{4N^2} (N+2) \left[ \langle \phi_0^2 \rangle_0^2 - \langle \phi_0^4 \rangle_0 \right] V_d [\hat{G}^{(m)}]_{ren} \notag \\
 &&\,\,\,\,\,\,\,\,\,\,\,\,\,+ \frac{\lambda^2}{32N^3} (N+2)^2 \left[ \langle \phi_0^6 \rangle_0 - 3 \langle \phi_0^2 \rangle_0 \langle \phi_0^4 \rangle_0 + 2 \langle \phi_0^2 \rangle_0^3 \right] V_d^2 [\hat{G}^{(m)}]_{ren}^2 \notag \\
 &&\,\,\,\,\,\,\,\,\,\,\,\,\,- \frac{\lambda^2}{16N^3} (N+8) \left[ \langle \phi_0^6 \rangle_0 - \langle \phi_0^2 \rangle_0 \langle \phi_0^4 \rangle_0 \right] V_d \left( \frac{\partial [\hat{G}^{(m)}]}{\partial m^2}\right)_{fin} \Biggr\} \notag \\
 &=& \frac{\delta_{ab}}{V_d m_{dyn}^2(IR)}. \label{2-pt-IR-ren}
\end{eqnarray}
The last equality follows after  interpreting the corrections as   a modification to the mass $m_{dyn}^2(IR)$ of the zero modes which, as mentioned before, determines the curvature of the effective potential.  

Eqs.~\eqref{2-pt-UV-ren} and \eqref{2-pt-IR-ren} are the main results of this section. They contain the main corrections to the renormalized UV and IR propagators, for any values of $d$ and $N$.

\section{Massless fields}\label{subsec-massless}

A case of great interest is when the fields are massless at tree level, $m=0$, as it is in this case in which the perturbative treatment becomes ill-defined. The nonperturbative treatment of the zero modes ensures that these modes acquire a dynamical mass, avoiding the  IR divergence associated to the free two-point functions 
in the massless limit. This can be verified by checking that Eq.~\eqref{2-pt-IR-ren} remains finite in this limit. Indeed, the $n$-point functions of the zero modes can be exactly computed to be
\begin{eqnarray}
 \langle \phi_0^{2p} \rangle_0 = \frac{\int_0^{\infty} d\phi_0 \, \phi_0^{N-1+2p} e^{-\frac{V_d \lambda}{8N} \phi_0^4}}{\int_0^{\infty} d\phi_0 \, \phi_0^{N-1} \, e^{-\frac{V_d \lambda}{8N} \phi_0^4}}
  =2^{\frac{3p}{2}} \left( \frac{N}{V_d \lambda} \right)^{\frac{p}{2}} \frac{\Gamma\left[ \frac{N+2p}{4} \right]}{\Gamma\left[ \frac{N}{4} \right]}, \label{phi_0-2p-massless}
\end{eqnarray}
which exhibit no IR divergences. This equation shows a scaling of the form $\phi_0 \sim \lambda^{-1/4}$, making the perturbative expansion of the UV modes to be in powers of $\sqrt{\lambda}$. 

It is worth to note that, when considering the two-point functions of the UV modes, Eq.~\eqref{2-pt-UV-ren}, the free UV propagators that build up this expression will now become massless. After performing the analytical continuation to the Lorentzian spacetime, this leads to an IR enhanced behavior at large distances, i.e. secular growth (see Appendix \ref{app-UVprop}). The reason for this is that so far we have only given mass to the zero modes, while the UV modes remain massless, as sketched diagramatically in Appendix \ref{app-Double-expansion}. This situation can be dealt with by improving the result Eq.~\eqref{2-pt-UV-ren} by further resumming a given subset of diagrams that give mass to the UV propagators appearing
in that expression. We will focus on this point in  Section \ref{Ressum}.

\subsection{The 1/N expansion}

In  Euclidean de Sitter space, after dealing with the zero modes, the calculations of the $n$-point functions can be done by means of a perturbative expansion in powers of $\sqrt{\lambda}$.  When  $\lambda$ is sufficiently small,  being a compact space, this perturbative expansion is valid for any set of points and for any value of $N$. So an expansion in $1/N$ is certainly unnecessary in that case. However, we are ultimately interested in computing quantities in the Lorentzian de Sitter spacetime.  For this,  as  will become clear in the next section,  a double expansion in $\sqrt{\lambda}$ and $1/N$ turns out to be crucially convenient in order to obtain a tractable perturbative expansion that remains valid at large distances.
With this in mind, we  perform an expansion  in $1/N$ of  the  results obtained in the previous section. In order to compare with  known  results appearing in the literature, it is sufficient to remain at NLO in $1/N$. At the same time, the  known Lorentzian results that are nonperturbative in the coupling constant will have to be expanded in powers of $\sqrt{\lambda}$ to bring them to the same precision.

The Euclidean results for the two-point functions at order $\lambda$ and order $1/N$ are obtained by inserting Eq.~\eqref{phi_0-2p-massless}  into Eqs.~\eqref{2-pt-UV-ren} and \eqref{2-pt-IR-ren},  and then expanding in powers of $1/N$. We arrive at the following expressions:
\begin{eqnarray}
  \langle \hat{\phi}_{a}(x) \hat{\phi}_{b}(x') \rangle &=& \delta_{ab} \Biggl\{ \hat{G}^{(0)}(x,x') + \left( \sqrt{\frac{\lambda}{2V_d}} + \frac{\lambda}{4} [\hat{G}^{(0)}]_{ren} \right) \frac{\partial \hat{G}^{(m)}(x,x')}{\partial m^2} \Bigg|_0 \notag \\
  &&+ \frac{1}{2} \frac{\lambda}{2V_d} \frac{\partial^2 \hat{G}^{(m)}(x,x')}{\partial (m^2)^2} \Bigg|_0 
  + \frac{1}{2N} \Biggl[ \left( 3 \sqrt{\frac{\lambda}{2V_d}} + \frac{\lambda}{4} [\hat{G}^{(0)}]_{ren} \right) \frac{\partial \hat{G}^{(m)}(x,x')}{\partial m^2} \Bigg|_0 \notag \\
  &&+ \frac{4\lambda}{V_d} \frac{\partial^2 \hat{G}^{(m)}(x,x')}{\partial (m^2)^2} \Bigg|_0 \Biggr] \Biggr\}, \label{2-pt-UV-massless-NLO-N-1-expanded}
\end{eqnarray}
and
\begin{eqnarray}
 \langle \phi_{0a} \phi_{0b} \rangle& =& \delta_{ab} \Biggl\{  \sqrt{\frac{2}{V_d\lambda}} - \frac{1}{2} [\hat{G}^{(0)}]_{ren} + \frac{1}{8} \sqrt{\frac{V_d \lambda}{2}} [\hat{G}^{(0)}]_{ren}^2 \notag - \frac{1}{2} \sqrt{\frac{\lambda}{2V_d}} \left( \frac{\partial [\hat{G}^{(m)}]}{\partial m^2} \right)_{0,fin} \notag \\
 &&+ \frac{1}{2N} \Biggl[ - \sqrt{\frac{2}{V_d\lambda}} - \frac{3}{2} [\hat{G}^{(0)}]_{ren} + \frac{9}{8} \sqrt{\frac{V_d \lambda}{2}} [\hat{G}^{(0)}]_{ren}^2 + \frac{15}{2} \sqrt{\frac{\lambda}{2V_d}} \left( \frac{\partial [\hat{G}^{(m)}]}{\partial m^2} \right)_{0,fin}
 \Biggr] \Biggr\}, 
 \notag\\ &&
  \label{2-pt-IR-massless-largeN} 
\end{eqnarray}
for the UV and IR contributions, respectively.  It is worth to stress that the leading contribution in the limit $N \to \infty$ of Eq.~\eqref{2-pt-UV-massless-NLO-N-1-expanded} is compatible with a Taylor expansion of a massive  UV propagator, with an  UV dynamical mass given by 
\begin{equation}
m_{dyn}^2(UV) = \sqrt{\frac{\lambda}{2V_d}} + \frac{\lambda}{4} [\hat{G}^{(0)}]_{ren},
\end{equation}
up to order $\lambda$.
 Furthermore, the IR dynamical mass $m_{dyn}^2(IR)$ read from Eq.~\eqref{2-pt-IR-massless-largeN} also coincides up to that order with $m_{dyn}^2(UV)$, and therefore the whole propagator can be interpreted as a free de Sitter propagator with a dynamical mass $m_{dyn}^2$. 
Beyond the LO contribution in $1/N$, this is not true, since in Eq.~\eqref{2-pt-UV-massless-NLO-N-1-expanded} the coefficient of the second derivative is no longer half the square of that of the first derivative, and the two-point functions 
has a more complicated structure.

\subsection{Comparison with Lorentzian QFT: dynamical mass}

There are several nonperturbative approaches in the Lorentzian QFT. Here we consider the 2PIEA formulation, in which both the mean value of the field $\bar{\phi}(x)$ and the exact two-point functions 
$G^{(m_{dyn})}(x,x')$ are treated as independent degrees of freedom. A detailed description of this method is given elsewhere. In this framework it is possible to obtain the exact result for the two-point functions in the large-$N$ limit. For this, we follow \cite{Hartree2,JPP} and write down the exact equations of motion in the large-$N$ limit, 
\begin{eqnarray}
 \Biggl[ -\square + m^2 + \frac{\lambda}{2} \bar{\phi}^2 + \frac{\lambda}{2} [G^{(m_{dyn})}] \Biggr] \bar{\phi} &=& 0, \label{bar-phi} \\
 \Biggl[ -\square + m^2 + \frac{\lambda}{2} \bar{\phi}^2 + \frac{\lambda}{2} [G^{(m_{dyn})}] \Biggr] G^{(m_{dyn})}(x,x') &=& i \frac{\delta(x-x')}{\sqrt{-g}}. \label{prop-eq-2PI} 
\end{eqnarray}
In this limit, the equations become local. Eq.~\eqref{prop-eq-2PI} corresponds to that of a free propagator in de Sitter spacetime  with  mass 
$m_{dyn}^2$ satisfying a self-consistent gap equation,
\begin{equation}
 m_{dyn}^2 = m^2 + \delta m^2 + \frac{(\lambda+\delta\lambda)}{2} \bar{\phi}^2 + \frac{(\lambda+\delta\lambda)}{2} [G^{(m_{dyn})}],
\end{equation}
where the counterterms have to be suitably chosen to cancel the divergences of the coincidence limit of the propagator in the right hand side. We expand $[G^{(m_{dyn})}]$ in powers of $m_{dyn}^2$, 
\begin{equation}
[G^{(m_{dyn})}] = \frac{1}{V_d m_{dyn}^2} + [\hat{G}^{(0)}] + m_{dyn}^2 \left( \frac{\partial [\hat{G}^{(m)}]}{\partial m^2} \right)_0 + \dots \label{exact-G-mdyn-exp}
\end{equation}
where the dots stand for terms of order $\mathcal{O}(m_{dyn}^4/H^4)$ and can be neglected since we are interested in the small mass case $m_{dyn}^2 \ll H^2$. We take advantage of the fact that the coincident propagator is exactly the same for both the Lorentzian and Euclidean theories. Therefore we use the same notation as for our Euclidean calculations for the different parts in the expansion. The necessary counterterms are
\begin{eqnarray}
 \delta m^2 &=& - \frac{\lambda}{2} \left[ \frac{[\hat{G}^{(0)}]_{div} + m^2 \left( \frac{\partial [\hat{G}^{(m)}]}{\partial m^2} \right)_{0,div} }{ 1 + \frac{\lambda}{2} \left( \frac{\partial [\hat{G}^{(m)}]}{\partial m^2} \right)_{0,div} } \right], \\
 \delta \lambda &=& - \frac{\frac{\lambda^2}{2} \left( \frac{\partial [\hat{G}^{(m)}]}{\partial m^2} \right)_{0,div} }{\left[ 1 + \frac{\lambda}{2} \left( \frac{\partial [\hat{G}^{(m)}]}{\partial m^2} \right)_{0,div} \right]}.
\end{eqnarray}
When expanded at the leading order in $\lambda$, these counterterms coincide with those used in the Euclidean calculation, Eqs.~\eqref{delta-m-euclidean} and \eqref{delta-lambda-euclidean}, when the latter are expanded for small masses and the large-$N$ limit is taken. The renormalized gap equation is then, in the symmetric phase $\bar{\phi} = 0$:
\begin{eqnarray}
m_{dyn}^2 = m^2 + \frac{\lambda}{2} \Biggl[ \frac{1}{V_d m_{dyn}^2} + [\hat{G}^{(0)}]_{ren} + m_{dyn}^2 \left( \frac{\partial [\hat{G}^{(m)}]}{\partial m^2} \right)_{0,fin} \Biggr]. 
\end{eqnarray}
This is a quadratic algebraic equation for $m_{dyn}^2$, whose positive and physically relevant solution is
\begin{equation}
 m_{dyn}^2 = \frac{-b + \sqrt{ b^2 + \frac{2\lambda}{V_d} a}}{2 a},
\end{equation}
where we have defined
\begin{eqnarray}
a &=& 1 - \frac{\lambda}{2} \left( \frac{\partial [\hat{G}^{(m)}]}{\partial m^2} \right)_{0,fin}, \\
b &=& m^2 + \frac{\lambda}{2} [\hat{G}^{(0)}]_{ren}.
\end{eqnarray}
The dynamical mass is finite for $m = 0$, and exact in the large-$N$ limit. 

In order to draw a comparison with the Euclidean results,
we expand the Lorentzian expression up to order $\sqrt{\lambda}$,
\begin{eqnarray}
 \frac{1}{V_d m_{dyn}^2} &=& \sqrt{\frac{2}{V_d\lambda}} - \frac{1}{2} [\hat{G}^{(0)}]_{ren} + \frac{1}{8} \sqrt{\frac{V_d \lambda}{2}} [\hat{G}^{(0)}]_{ren}^2 - \frac{1}{2} \sqrt{\frac{\lambda}{2V_d}} \left( \frac{\partial [\hat{G}^{(m)}]}{\partial m^2} \right)_{0,fin} . 
\end{eqnarray}
The corresponding Euclidean calculation is given in Eq.~\eqref{2-pt-IR-massless-largeN}. We see that, at leading order in $1/N$, the dynamical masses computed in both approaches coincide. 

When going beyond the leading order in $1/N$, the previous approach is no longer valid, since the propagator cannot be described as a free propagator with a dynamical mass given by the gap equation, Eq.~\eqref{prop-eq-2PI}, and the full Schwinger-Dyson equations must be solved instead. The complete Lorentzian results that take into account the UV modes and the renormalization process are technically involved and still unknown. However, in Ref. \cite{Gautier} the authors were able to obtain results up to the NLO in the $1/N$ expansion which are valid at the leading order in the IR. We refer the reader to their paper for the details. Basically, exploiting the de Sitter symmetries, the scalar fields are split into two sectors: an IR one, formed by modes with physical momenta smaller than a critical scale $\mu_{IR}\ll H$, and the one containing the remaining modes. The interactions of the IR modes are taken into account, but the ones of the remaining modes are neglected. In this approximation, the authors have obtained a self-consistent solution of the  Schwinger-Dyson  equations for the IR two-point functions that is valid at NLO in $1/N$ and at leading order in the IR. In particular, the  result for the dynamical mass can be written as \cite{Gautier}:
  \begin{equation}
  \frac{1}{V_d m_{dyn}^2}= \sqrt{\frac{2}{V_d\lambda}} \left(1-\frac{1}{2N}\right)+...,
 \end{equation} where the dots stand for corrections that are higher order in either $\sqrt{\lambda}$ or $1/N$ (which cannot be unambiguously computed  in the approximation considered, since they depend on the arbitrary IR scale $\mu_{IR}$).  Clearly, this result agrees with our Eq.~\eqref{2-pt-IR-massless-largeN} at  leading order in $\sqrt{\lambda}$.

\section{Resumming the leading IR secular terms to the two-point functions}\label{Ressum}

In section \ref{subsec-massless} we discussed the case of massless fields and took notice that the calculated two-point functions 
of the UV modes, given in Eq.~\eqref{2-pt-UV-ren},  would be expressed  in terms of massless UV propagators. The problem is then that the analytically continued  correlation function does not  decay at large distances, as happens  for massive fields. The reason for this being that the UV modes remained massless in this framework. 

In this  section we will extend the nonperturbative treatment, with the aim of resumming the leading IR secular terms. In order to achieve this, we need to perform a resummation of diagrams to give mass to the UV propagators present in Eq.~\eqref{2-pt-UV-ren}. As we will see,  it will be enough to resum only a subclass of diagrams: those coming from the interaction term that is quadratic in both $\phi_0$ and $\hat{\phi}$,
\begin{equation}\label{biq}
 S^{(2)}_{int}[\phi_0,\hat{\phi}] = \frac{\lambda}{4N} \int d^d x \sqrt{g} \, A_{abcd} \phi_{0a} \phi_{0b} \hat{\phi}_c \hat{\phi}_d, 
\end{equation}
which will be treated nonperturbatively, while still perturbing on the remaining terms.

We start by rewritting the generating functional Eq.~\eqref{Z-int} by grouping this term with the other terms quadratic in $\hat{\phi}$, as part of the ``free'' generating functional of the UV modes,
\begin{eqnarray}
Z[J_0, \hat{J}] &=& \mathcal{N} e^{-\tilde{\tilde{S}}_{int}\left[\frac{\delta}{\delta J_0},\frac{\delta}{\delta \hat{J}}\right]} \int d^N \phi_0 \, e^{-\left[ \frac{\lambda V_d}{8N} |\phi_0|^4 + V_d J_{0a} \phi_{0a} \right]} \notag \\
&&\times \int \mathcal{D}\hat{\phi} \, \exp\left( -\frac{1}{2} \iint_{x,y} \hat{\phi}_a  \hat{G}^{-1}_{ab}(\phi_0) \hat{\phi}_b  + \int_x \hat{J}_a \hat{\phi}_a \right) \notag \\
&=& \mathcal{N} e^{-\tilde{\tilde{S}}_{int}\left[\frac{\delta}{\delta J_0},\frac{\delta}{\delta \hat{J}}\right]} \left( \hat{Z}_{f}\left[\hat{J}, m^2 \right] \left[ \det  \hat{G}_{rs}^{(m)} \right]^{1/2} \right)_{m(\delta/\delta J_0)} Z_0[J_0],
\end{eqnarray}
where now the ``free'' UV propagator $\hat{G}_{ab}(\phi_0)$ has a $\phi_0$-dependent mass,
\begin{equation}
 \hat{G}^{-1}_{ab}(\phi_0)(x,x') = \left[-\square \delta_{ab} + m_{ab}^2(\phi_0) \right] \frac{ \delta^{(d)}(x-x')}{\sqrt{g}},
 \label{inv-hatG}
\end{equation}
with $m_{ab}^2(\phi_0) = (\lambda/2N) A_{abcd} \phi_{0c} \phi_{0d}$ and where $\tilde{\tilde{S}}_{int} = \tilde{S}_{int} - S^{(2)}_{int}$ has the remaining interaction terms that should be treated perturbatively. The normalization factor $\mathcal{N}$ ensures that $Z[0, 0]=1$.

In order to compare with the results of the previous sections, it will be enough to keep terms up to order $\lambda$. Therefore,
it is necessary to include perturbatively only the first correction coming from the term, 
\begin{equation}
 S^{(4)}_{int}[\hat{\phi}] = \frac{\lambda}{8N} \int_x |\hat{\phi}|^4,
\end{equation}
and so the generating functional reduces to
\begin{eqnarray}
 Z^{(1)}[J_0, \hat{J}] &=& \left[ 1 - \frac{\lambda}{8N} \delta_{cd} \delta_{ef} \int_x \frac{\delta^4}{\delta \hat{J}^4_{cdef}(x)} \right] \left( \hat{Z}_{f}\left[\hat{J}, m^2 \right] \left[ \det  \hat{G}_{rs}^{(m)} \right]^{1/2} \right)_{m\left(\frac{\delta}{\delta J_0}\right)} Z_0[J_0]. \,\,\,\,\,\,\,\,\,\,\,\, \label{eq5.5}
\end{eqnarray} 
%
Now we proceed to calculate the connected two-point functions 
of the UV modes,
\begin{eqnarray}
 \langle \hat{\phi}_{a}(x) \hat{\phi}_{b}(x') \rangle^{(1)} &=& \frac{1}{Z^{(1)}[0, 0]} \frac{ \delta^2 Z^{(1)}[J_0,\hat{J}] }{\delta \hat{J}_a(x) \delta \hat{J}_b(x')} \Bigg|_{J_0,\hat{J}=0} \label{eq5.6}\\
 &=& \langle \hat{\phi}_{a}(x) \hat{\phi}_{b}(x') \rangle^{(0)} + \Delta \langle \hat{\phi}_{a}(x) \hat{\phi}_{b}(x') \rangle, \notag
\end{eqnarray}
which we split in two contributions, according to the interaction term that we are treating perturbatively. Approximating $Z^{(1)}$ in Eq.~\eqref{eq5.6} by the first term of Eq.~\eqref{eq5.5} we obtain for the first contribution, 
\begin{eqnarray}
  \langle \hat{\phi}_{a}(x) \hat{\phi}_{b}(x') \rangle^{(0)} &=& 
  \frac{\left( \hat{G}^{(m)}_{ab}(x,x') \left[ \det  \hat{G}_{rs}^{(m)} \right]^{1/2} \right)_{m\left(\frac{\delta}{\delta J_0}\right)} Z_0[J_0] \Bigg|_{J_0=0} }{\left[ \det \hat{G}_{rs}^{(m(\delta/\delta J_0))} \right]^{1/2} Z_0[J_0] \Big|_{J_0=0}}. \label{2-pt-UV-resum}
\end{eqnarray}
Here it is important to note that both square roots of the determinant of the propagator will not cancel each other out due to the integral over $\phi_0$. This is crucial to obtain the correct result.
In order to evaluate this formal expression, we must expand both the numerator and the denominator in powers of $m_{ab}^2(\delta/\delta J_0)$. We start with the numerator:
\begin{eqnarray}
  \sum_{p=0}^{\infty} \frac{1}{p!} \frac{\partial^p \left(\hat{G}_{ab}^{(m)}(x,x') \left[ \det \hat{G}_{rs}^{(m)} \right]^{1/2} \right)}{\partial m_{i_1 j_1}^2 \dots \partial m_{i_p j_p}^2} \Bigg|_{0} \left( \prod_{\alpha = 1}^p \frac{\lambda}{2N} A_{i_\alpha j_\alpha k_\alpha l_\alpha} \right) \frac{\delta^{2p} Z_0[J_0]}{\delta J_{0k_1} \delta J_{0l_1} \dots \delta J_{0k_p} \delta J_{0l_p}} \Bigg|_{J_0=0}.
\end{eqnarray}
The key point is that our resummation needs only include the (infinite) subset of contributions that modify the UV propagator at separate points, while the determinant of  $\hat{G}_{ab}$ has no IR problems and can be safely evaluated at $m=0$. Therefore, in the above series it is enough to consider only terms with as many derivatives acting on the determinant as needed for a given precision in $\lambda$, since each time we increase the number of those derivatives we pick up a factor of $\lambda \, \phi_0^2 \sim \sqrt{\lambda}$. In our case, we shall keep terms with zero and one derivatives of the determinant. 

\begin{figure}[htbp]
\begin{center}
\begin{fmfgraph*}(120,100)
\fmfstraight
\fmfleft{i} \fmfright{o} \fmftop{t1,t2,t3} \fmf{plain,tension=100}{i,v0,v1,v2,v3,v4,v5,v6,o} \fmfpolyn{smooth,label=$\langle \phi_0^{2p} \rangle_0$,tension=2,filled=20}{p}{20} \fmf{dashes,tension=0.05}{p1,v1,p2} \fmf{dashes,tension=0.01}{p3,v2,p4}
\fmf{phantom,tension=0.01}{p5,v3,p6} \fmf{phantom,tension=0.01}{p7,v4,p8} \fmf{dashes,tension=0.05}{p9,v5,p10} \fmf{phantom,tension=1}{p18,t1} \fmf{phantom,tension=1}{p13,t3} 
\fmfv{l=$x$,l.a=180}{i}
\fmfv{l=$x'$,l.a=0}{o}
\fmfv{l=$1$,l.a=-90}{v1}
\fmfv{l=$2$,l.a=-90}{v2}
\fmfv{l=$\dots$,l.a=-60,l.d=8}{v3}
\fmfv{l=$p$,l.a=-90}{v5}
\fmfdot{v1} \fmfdot{v2} \fmfdot{v5}
\end{fmfgraph*}
\vspace{-1.3cm}
\caption{This diagram represents a particular term in the series of Eq.~\eqref{series-det} with $p$ insertions of $(\lambda/2N) A_{ijkl} \phi_{0j} \phi_{0k}$. The solid lines stand for UV propagators $\hat{G}$ while the dashed lines represent $\phi_0$'s. The solid blob indicates a nonperturbative correlation function of zero modes obtained from $Z_0[J_0]$.}
\label{Fig1}
\end{center}
\end{figure}
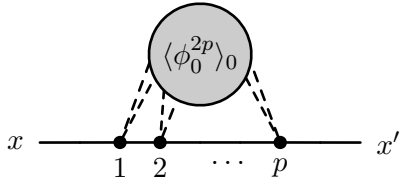

\begin{figure}[htbp]
\begin{center}
\begin{fmfgraph*}(120,50)
\fmfstraight
\fmfleft{i,im,it} \fmfright{o,om,ot} \fmftop{t1,t2,t3} \fmf{plain,tension=90}{i,v0,v1,v2,v3,v4,v5,v6,o} \fmfpolyn{smooth,label=$\langle \phi_0^{2p} \rangle_0$,tension=2,filled=20}{p}{20} \fmf{dashes,tension=0.05}{p1,v1,p2} \fmf{dashes,tension=0.01}{p3,v2,p4}
\fmf{phantom,tension=0.01}{p5,v3,p6} \fmf{dashes,tension=0.01}{p7,v4,p8} \fmf{phantom,tension=0.05}{p9,v5,p10} \fmf{phantom,tension=1}{p18,t1} \fmf{phantom,tension=1}{p13,t3} 
\fmf{dashes,tension=0.1}{p10,v7,p12}
\fmf{plain,left,tension=0.05}{v7,om,v7}
\fmfv{l=$x$,l.a=180}{i}
\fmfv{l=$x'$,l.a=0}{o}
\fmfv{l=$1$,l.a=-90}{v1}
\fmfv{l=$2$,l.a=-90}{v2}
\fmfv{l=$\dots$,l.a=-100,l.d=8}{v3}
\fmfv{l=$p-1$,l.a=-120,l.d=6}{v5}
\fmfdot{v1} \fmfdot{v2} \fmfdot{v4} \fmfdot{v7}
\end{fmfgraph*}
\vspace{0.5cm}
\caption{Similar to Fig. \ref{Fig1} but in this case representing a term in the series of Eq.~\eqref{series-deriv-det} instead. The loop of UV modes comes from the derivative of the determinant, and it is disconnected from the other solid lines (although it is connected through the interactions with the zero modes).}
\label{Fig2}
\end{center}
\end{figure}
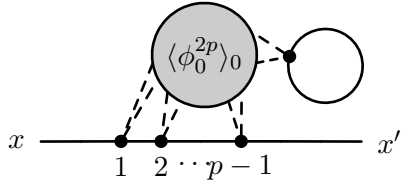

If first we consider the subset of terms with no derivatives of the determinant, we have a series for the connected UV propagator,
\begin{eqnarray}
 \left[ \det \hat{G}_{rs}^{(0)} \right]^{1/2} \sum_{p=0}^{\infty} \frac{1}{p!} \frac{\partial^p \hat{G}_{ab}^{(m)}(x,x')}{\partial m_{i_1 j_1}^2 \dots \partial m_{i_p j_p}^2} \Bigg|_{0} \left( \prod_{\alpha = 1}^p \frac{\lambda}{2N} A_{i_\alpha j_\alpha k_\alpha l_\alpha} \right) \langle \phi_{0 k_1} \phi_{0 l_1} \dots \phi_{0 k_p} \phi_{0 l_p} \rangle_0. \label{series-det}
\end{eqnarray}
We see that each term of the series corresponds to a connected diagram with $p$ insertions of $(\lambda/2N) A_{ijkl} \phi_{0j} \phi_{0k}$ and two external legs, as shown in Fig. \ref{Fig1}. The contractions of the $O(N)$ indices amount to an overall $\delta_{ab}$ and a factor depending both on $N$ and $p$. To calculate this factor, we recall that each $A_{ijkl}$ has 3 terms made of pairs of Kronecker deltas, so for each $p$ there will be $3^p$ terms. Of these, one term will be proportional to
\begin{equation}
 \delta_{ab} \langle \phi_0^{2p} \rangle_0,
\end{equation}
while the other $3^p - 1$ terms will leave two powers of $\phi_0$ untraced,
\begin{equation}
 \langle \phi_{0a} \phi_{0b} \, \phi_0^{2(p-1)} \rangle_0 = \frac{\delta_{ab}}{N} \langle \phi_0^{2p} \rangle_0.
\end{equation}
Therefore, the series in Eq.~\eqref{series-det} becomes
\begin{fmffile}{diagram1}
\fmfset{dot_len}{0.7mm}
\end{fmffile}
\begin{eqnarray}
 \left[ \det \hat{G}_{rs}^{(0)} \right]^{1/2} \delta_{ab} \sum_{p=0}^{\infty} \frac{1}{p!} \frac{\partial^p \hat{G}^{(m)}(x,x')}{\partial (m^2)^p} \Bigg|_{0} \left( \frac{\lambda}{2N} \right)^p \left[ 1 + \frac{(3^p-1)}{N} \right] \langle \phi_0^{2p} \rangle. 
\end{eqnarray}
This series can be resummed order by order in $1/N$, as we will see in the next section. Before doing this, let us first deal with the other subset of terms that we need to consider up to order $\lambda^1$, namely, those in which there is one derivative acting on the determinant,
\begin{eqnarray}
 &&\frac{\partial \left[ \det \hat{G}_{rs}^{(m)} \right]^{1/2}}{\partial m_{i j}^2} \Bigg|_{0} \frac{\lambda}{2N} A_{ijkl} \label{series-deriv-det} \\
 && \times \sum_{p=0}^{\infty} \frac{p}{p!}  \frac{\partial^{p-1} \hat{G}_{ab}^{(m)}(x,x')}{\partial m_{i_1 j_1}^2 \dots \partial m_{i_{p-1} j_{p-1}}^2} \Bigg|_{0} \left( \prod_{\alpha = 1}^{p-1} \frac{\lambda}{2N} A_{i_\alpha j_\alpha k_\alpha l_\alpha} \right) \langle \phi_{0 k} \phi_{0 l} \phi_{0 k_1} \phi_{0 l_1} \dots \phi_{0 k_{p-1}} \phi_{0 l_{p-1}} \rangle_0. \notag
\end{eqnarray}
In this case the UV propagators in the diagrams are not all connected among themselves, but rather through their interaction with the zero modes. Indeed, these diagrams are composed of a single bubble with one insertion of $(\lambda/2N) A_{ijkl} \phi_{0j} \phi_{0k}$ (factorized in the previous expression) times a connected part with $p-1$ insertions, as depicted in Fig. \ref{Fig2}. The prefactor is simply proportional to $\delta_{kl}$, while the series now contains only connected diagrams, and the same argument as before applies. Therefore, relabeling the summation index $p=l+1$, we obtain
\begin{eqnarray}
 \frac{\partial \left[ \det \hat{G}_{rs}^{(m)} \right]^{1/2}}{\partial m^2} \Bigg|_{0} \frac{\lambda(N+2)}{2N} \delta_{ab} \sum_{l=0}^{\infty} \frac{1}{l!}  \frac{\partial^l \hat{G}^{(m)}(x,x')}{\partial (m^2)^l} \Bigg|_{0} \left( \frac{\lambda}{2N} \right)^l \left[ 1 + \frac{(3^l-1)}{N} \right] \langle \phi_0^{2(l+1)} \rangle_0. \,\,\,\,\,\,\,\,\,\,
\end{eqnarray}

\begin{fmffile}{diagram2}
\fmfset{dot_len}{0.7mm}
\end{fmffile}

Next we consider the denominator of Eq.~\eqref{2-pt-UV-resum}. As we discussed, there is no need to resum the determinant since it has no external legs. Therefore we can treat it  perturbatively in $\lambda$. To be consistent with what we did with the numerator, we should keep the first two terms, that is, with zero and one derivatives acting on the determinant,
\begin{eqnarray}
 \frac{1}{\left[ \det \hat{G}_{rs}^{(m(\delta/\delta J_0))} \right]^{1/2} Z_0[J_0] \Big|_{J_0=0}} &=& \frac{1}{\left[ \det \hat{G}_{rs}^{(0)} \right]^{1/2}} \left[ 1 - \frac{\lambda(N+2)}{2N} \langle \phi_0^2 \rangle_0 \left( \frac{\frac{\partial \left[ \det \hat{G}_{rs}^{(0)} \right]^{1/2}}{\partial m^2}}{N \left[ \det \hat{G}_{rs}^{(0)} \right]^{1/2}} \right) \right] \notag \\
 &&+ \mathcal{O}(\lambda).
\end{eqnarray}
The combination in big parentheses can be easily computed (see Appendix \ref{app-deriv-det}) to give
\begin{equation}
 \left( \frac{\frac{\partial \left[ \det \hat{G}_{rs}^{(0)} \right]^{1/2}}{\partial m^2}}{N \left[ \det \hat{G}_{rs}^{(0)} \right]^{1/2}} \right) = - \frac{V_d}{2} [\hat{G}^{(0)}], \label{partial-det}
\end{equation}
where, although there is a massless propagator, there are no infrared problems since it is evaluated at its coincidence limit. To deal with the UV divergence, the renormalization can be carried out by introducing (at this order) the mass counterterm $\delta m^2$ given in Eq.~\eqref{delta-m-euclidean} at $m=0$, both in the action of the zero modes as well as for the UV modes. Then, when performing the above mass expansions, the diagrams now can be constructed with both lines with insertions of $(\lambda/2N) A_{ijkl} \phi_{0j} \phi_{0k}$ and of $\delta m^2 \delta_{ij}$. As we did with the derivatives of the determinant, we need only keep a finite number of $\delta m^2 \delta_{ij}$ insertions, given that $\delta m^2 \sim \lambda$. In our case, we need only one, but in general we will need as many as derivatives of the determinant we have. At higher orders in $\lambda$ it is also necessary to include a $\delta \lambda$ counterterm.

Putting everything together, distributing, and keeping the terms up to the corresponding order in $\lambda$, we arrive at
\begin{eqnarray}
  \langle \hat{\phi}_{a}(x) \hat{\phi}_{b}(x') \rangle^{(0)} &=& \delta_{ab} \Biggl\{ \sum_{p=0}^{\infty} \frac{1}{p!} \frac{\partial^p \hat{G}^{(m)}(x,x')}{\partial (m^2)^p} \Bigg|_{0} \left( \frac{\lambda}{2N} \right)^p \left[ 1 + \frac{(3^p-1)}{N} \right] \langle \phi_0^{2p} \rangle_0 \label{UV-prop-allN} \\
  &&- \frac{\lambda^2(N+2)}{4N} V_d [\hat{G}^{(0)}]_{ren} \times \notag \\
  && \sum_{l=0}^{\infty} \frac{1}{l!}  \frac{\partial^l \hat{G}^{(m)}(x,x')}{\partial (m^2)^l} \Bigg|_{0} \left( \frac{\lambda}{2N} \right)^l \left[ 1 + \frac{(3^l-1)}{N} \right] \left[ \langle \phi_0^{2(l+1)} \rangle_0 - \langle \phi_0^{2} \rangle_0 \langle \phi_0^{2l} \rangle_0 \right] \Biggr\}. \notag
\end{eqnarray}

Now we look at the remaining term coming form the perturbative correction,
\begin{eqnarray}
\Delta \langle \hat{\phi}_{a}(x) \hat{\phi}_{b}(x') \rangle = \delta_{ab} \frac{\lambda (N+2)}{2N} \frac{\Biggl( [\hat{G}^{(m)}] \frac{\partial \hat{G}^{(m)}(x,x')}{\partial m^2} \left[ \det \hat{G}_{rs}^{(m)} \right]^{1/2} \Biggr)_{m\left(\frac{\delta}{\delta J_0}\right)} Z_0[J_0] \Bigg|_{J_0=0}}{\left[ \det \hat{G}_{rs}^{(m(\delta/\delta J_0))} \right]^{1/2} Z_0[J_0] \Bigg|_{J_0=0}}. \,\,\,\,\,\,\,\,
\end{eqnarray}
Once again, we must expand this in powers of the mass in order to evaluate the $J_0$-derivatives acting on $Z_0[J_0]$ an resum order by order in $1/N$. As before, the only part we should be concerned with resumming is the first derivative of the UV propagator at separate points, while the other factors are well behaved for $m=0$. The renormalization of the coincidence limit comes from the inclusion of the $\delta m^2$ counterterm. 

In contrast with the previous calculation, we have no need to keep any terms with derivatives of the determinant, since they would only contribute at higher orders in $\lambda$. Then, keeping only the contributions with derivatives over $\frac{\partial \hat{G}^{(m)}(x,x')}{\partial m^2}$, we have 
\begin{eqnarray}
\Delta \langle \hat{\phi}_{a}(x) \hat{\phi}_{b}(x') \rangle &=& \delta_{ab} \frac{\lambda (N+2)}{2N} [\hat{G}^{(0)}]_{ren} \label{Delta_hatphi_hatphi-all-N} \\
&& \times \sum_{p=0}^{\infty} \frac{1}{p!} \frac{\partial^{p+1} \hat{G}^{(m)}(x,x')}{\partial (m^2)^{p+1}} \Bigg|_{0} \left( \frac{\lambda}{2N} \right)^p \left[ 1 + \frac{(3^p-1)}{N} \right] \langle \phi_0^{2p} \rangle, \notag
\end{eqnarray}
where once again the counting of the $N$-dependent factor relies on the fact that each term in the series corresponds to a connected diagram.

\subsection{$1/N$ expansion at NLO}

As we mentioned, the series above can be resummed order by order in $1/N$, for which we need to expand the summands of the various series. Up to order $N^{-1}$ we have
\begin{eqnarray}
 \left( \frac{\lambda}{2N} \right)^p \left[ 1 + \frac{(3^p-1)}{N} \right] \langle \phi_0^{2p} \rangle_0 = \left( \sqrt{\frac{\lambda}{2V_d}} \right)^p \left[ 1 + \frac{p(p-2)+2(3^p - 1)}{2N} + \mathcal{O}\left(\frac{1}{N^2}\right) \right], \,\,\,\,\,\,\,\,\,\, \label{1-over-N-connected}
\end{eqnarray}
and
\begin{eqnarray}
 \frac{\lambda(N+2)}{2N} \left( \frac{\lambda}{2N} \right)^p \left[ 1 + \frac{(3^p-1)}{N} \right] && \left[ \langle \phi_0^{2(p+1)} \rangle_0 - \langle \phi_0^{2} \rangle_0 \langle \phi_0^{2p} \rangle_0 \right] \\
 &&= \left( \sqrt{\frac{\lambda}{2V_d}} \right)^{p+1} p \left[ 1 + \frac{3 + 2 \times 3^p + p(p - 3)}{2N} \right]. \notag
\end{eqnarray}
Using the first of these we can resum the first Taylor series of Eq.~\eqref{UV-prop-allN}:
\begin{eqnarray}
 \hat{G}^{(m)}(x,x')\Big|_{m_{dyn,0}} + \frac{1}{2N} &\Biggl[& -m^2 \frac{\partial \hat{G}^{(m)}(x,x')}{\partial m^2} + m^4 \frac{\partial^2 \hat{G}^{(m)}(x,x')}{\partial (m^2)^2} \notag \\
 &&+ 2 \hat{G}^{(\sqrt{3} m)}(x,x') - 2 \hat{G}^{(m)}(x,x') \Biggr]_{m_{dyn,0}} + \mathcal{O}\left(\frac{1}{N^2}\right), 
\end{eqnarray} 
where $m_{dyn,0}^2 = \sqrt{\frac{\lambda}{2V_d}}$, and the second series:
\begin{eqnarray}
 &&m_{dyn,0}^4 \Biggl\{ \frac{\partial \hat{G}^{(m)}(x,x')}{\partial m^2} + \frac{1}{2N} \Biggl[ \frac{\partial \hat{G}^{(m)}(x,x')}{\partial m^2} + 6 \frac{\partial \hat{G}^{(\sqrt{3}m)}(x,x')}{\partial m^2} \Biggr] + \mathcal{O}\left(\frac{1}{N^2}\right) \Biggr\}_{m_{dyn,0}} \notag \\
 &&+ \mathcal{O}(\lambda^{3/2}).
\end{eqnarray} 
Inserting these expressions into Eq.~\eqref{UV-prop-allN} we get,
\begin{eqnarray}
  \langle \hat{\phi}_{a}(x) \hat{\phi}_{b}(x') \rangle^{(0)} = \delta_{ab} &\Biggl\{& \hat{G}^{(m)}(x,x') - \frac{\lambda}{4} [\hat{G}^{(0)}]_{ren} \frac{\partial \hat{G}^{(m)}(x,x')}{\partial m^2} \label{UV-prop-NLO-N-0} \\
  &&+ \frac{1}{2N} \Biggl[ 2 \hat{G}^{(\sqrt{3} m)}(x,x') - 2 \hat{G}^{(m)}(x,x') \notag \\
  &&- \sqrt{\frac{\lambda}{2V_d}} \frac{\partial \hat{G}^{(m)}(x,x')}{\partial m^2} + \frac{\lambda}{2V_d} \frac{\partial^2 \hat{G}^{(m)}(x,x')}{\partial (m^2)^2}  \notag \\
  &&- \frac{\lambda}{4} [\hat{G}^{(0)}]_{ren} \left( \frac{\partial \hat{G}^{(m)}(x,x')}{\partial m^2} + 6 \frac{\partial \hat{G}^{(\sqrt{3} m)}(x,x')}{\partial m^2} \right) \Biggr] \Biggr\}_{m_{dyn,0}}. \notag
\end{eqnarray}
All the UV propagators at separated points now have a mass squared of order $\sqrt{\lambda}$, giving them the correct behavior at large separations. A noteworthy observation is the presence of some propagators whose squared mass is three times that of the others, something which could not have been anticipated from the perturbative result. Also, the only instance of a massless UV propagator in the previous expression has its coincidence limit taken, and it is 
therefore just a finite constant factor with no IR issues.

Now we turn to the remaining part, Eq.~\eqref{Delta_hatphi_hatphi-all-N}, where the series can be resummed as done for Eq.~\eqref{UV-prop-allN}. However, since the contribution at NLO in $1/N$ is also higher order in $\lambda$ with respect to the LO one, we can just keep the latter,
\begin{eqnarray}
\Delta \langle \hat{\phi}_{a}(x) \hat{\phi}_{b}(x') \rangle \label{UV-prop-NLO-N-Delta} = \delta_{ab} \frac{\lambda}{2} \left( 1 + \frac{2}{N} \right) [\hat{G}^{(0)}]_{ren} \frac{\partial \hat{G}^{(m_{dyn,0}^2)}(x,x')}{\partial m^2}. 
\end{eqnarray}

The full result for the connected two-point functions 
of the UV modes up to order $\lambda$ and $N^{-1}$, with a partial resummation of the infinite subset of diagrams, is obtained by combining Eqs.~\eqref{UV-prop-NLO-N-0} and \eqref{UV-prop-NLO-N-Delta}:
\begin{eqnarray}
  \langle \hat{\phi}_{a}(x) \hat{\phi}_{b}(x') \rangle^{(1)} = \delta_{ab} &\Biggl\{& \hat{G}^{(m)}(x,x') + \frac{\lambda}{4} [\hat{G}^{(0)}]_{ren} \frac{\partial \hat{G}^{(m)}(x,x')}{\partial m^2} \label{UV-prop-NLO-N-1} \\
  &&+ \frac{1}{2N} \Biggl[ 2 \hat{G}^{(\sqrt{3} m)}(x,x') - 2 \hat{G}^{(m)}(x,x') \notag \\
  &&- \sqrt{\frac{\lambda}{2V_d}} \frac{\partial \hat{G}^{(m)}(x,x')}{\partial m^2} +\frac{\lambda}{2V_d} \frac{\partial^2 \hat{G}^{(m)}(x,x')}{\partial (m^2)^2} \notag \\
  &&+ \frac{\lambda}{4} [\hat{G}^{(0)}]_{ren} \left(7 \frac{\partial \hat{G}^{(m)}(x,x')}{\partial m^2} - 6 \frac{\partial \hat{G}^{(\sqrt{3} m)}(x,x')}{\partial m^2} \right) \Biggr] \Biggr\}_{m_{dyn,0}}. \notag
\end{eqnarray}
It was verified as a cross-check that this result reduces to the perturbative one Eq.~\eqref{2-pt-UV-ren}, upon expanding the latter at NLO in $1/N$ (see Appendix \ref{app-resum-exp}).



The large distance behaviour of the two-point functions 
ultimately depends on the masses of the free propagators that build up the expression, $m_{dyn,0}^2$ and $3 m_{dyn,0}^2$, which determine how fast it decays. However, the true exponent could be other, the difference lying beyond the precision of this result.

\subsection{Comparison with Lorentzian QFT: two-point functions}

The Lorentzian calculations of Ref.~\cite{Gautier} for the two-point functions 
are expressed differently, as a sum of two free (full) propagators, 
\begin{equation}
 \langle \phi_a(x) \phi_b(x') \rangle =\delta_{ab}\left( c_+\, G^{(m_+)}(x,x') + c_-\, G^{(m_-)}(x,x')\right),
 \label{2-pt-Serreau}
\end{equation}
with masses 
\begin{eqnarray}
 m_+^2 &=& m_{dyn,0}^2 \left( 1 + \frac{1}{4N} \right),\\
 m_-^2 &=& 5\, m_{dyn,0}^2 \left( 1 + \frac{1}{4N} \right),
\end{eqnarray}
which are different to ours. The coefficients are given by $c_- = 5/16N$ and $c_+ + c_- = 1$. Therefore, in order to draw a comparison between both results, they are better expressed in terms of UV propagators and its derivatives around a common mass parameter, which we choose to be $m_{dyn,0}^2$. Also, the full propagators of Eq.~\eqref{2-pt-Serreau} must be first separated into their IR (constant) and UV parts, for which it is better to think of that expression as being analytically continued to Euclidean space.

Expanding the Euclidean result Eq.~\eqref{UV-prop-NLO-N-1} in this way and dropping higher order terms we obtain
\begin{eqnarray}
  \langle \hat{\phi}_{a}(x) \hat{\phi}_{b}(x') \rangle^{(1)} = \delta_{ab} &\Biggl\{& \hat{G}^{(m)}(x,x') + \frac{\lambda}{4} [\hat{G}^{(0)}]_{ren} \frac{\partial \hat{G}^{(m)}(x,x')}{\partial m^2} \notag \\
  &&+ \frac{1}{2N} \Biggl[ 3 \sqrt{\frac{\lambda}{2V_d}} \frac{\partial \hat{G}^{(m)}(x,x')}{\partial m^2} + 5 \frac{\lambda}{2V_d} \frac{\partial^2 \hat{G}^{(m)}(x,x')}{\partial (m^2)^2} \notag \\
  &&+ \frac{\lambda}{4} [\hat{G}^{(0)}]_{ren} \frac{\partial \hat{G}^{(m)}(x,x')}{\partial m^2} \Biggr] \Biggr\}_{m_{dyn,0}}, \label{UV-prop-NLO-N-1-bis}
\end{eqnarray}
while doing the same with the inhomogeneous part of the Lorentzian result Eq.~\eqref{2-pt-Serreau} yields all but those terms proportional to $[\hat{G}^{(m)}]_{ren}$. However, it is expected for these contributions to be missing in the Lorentzian result of Ref.~\cite{Gautier}, given the nonsystematic way the interactions among IR and UV sectors are treated there. Indeed, we can only trust that result up to the leading IR order, $\sqrt{\lambda}$, and up to NLO in $1/N$, in which case both results are compatible.

The convenience of the Euclidean approach becomes evident as it provides a systematic, order by order expansion in both $\sqrt{\lambda}$ and $1/N$, while also allowing for the inclusion of partial resummations that cure the IR effects at large separations.

\section{Conclusions}

In this paper we considered an interacting $O(N)$ scalar  field model in $d$-dimensional Euclidean de Sitter spacetime, paying particular attention to the IR problems that appear for massless and light fields. We extended the approach of Refs.~\cite{Rajaraman1,BenekeMoch} to the $O(N)$ model. The zero modes are treated exactly while the corrections due to the interactions with the UV modes are computed perturbatively. The calculation of the two-point functions of the field shows that the exact treatment of the zero modes cures the IR divergences of the usual massless propagator: the two-point functions becomes de Sitter invariant. However, the NLO contains derivatives of the free propagator of  the UV modes. Although the massless UV propagator is de Sitter invariant, its Lorentzian counterpart exhibits a growing behavior at large distances, invalidating the perturbative expansion in this limit. This problem can be fixed in the leading order large $N$ limit by resumming the higher order corrections: one can show that the final result corresponds to a two-point functions 
of a free field with a self-consistent mass. 
 
In order to alleviate the behavior of the correlation functions in the IR limit, we performed a resummation of a class of diagrams that give mass to the UV propagator. After this resummation, higher order corrections can be systematically computed in a perturbative expansion in powers of both $\sqrt\lambda$ and $1/N$. We presented explicit results up to second order in $\sqrt{\lambda}$ and NLO in $1/N$.
  
In the leading large $N$ limit, the results can be computed exactly, both in the Euclidean approach and also working directly in the Lorentzian spacetime (which corresponds to the Hartree approximation \cite{Hartree1,Hartree2}). We  derived the corresponding results in the Lorentzian spacetime paying particular attention to the renormalization process. We showed that the Euclidean approach reproduces the Lorentzian results in this large $N$ limit, provided the same renormalization scheme is used. 
As we mentioned before, to our knowledge, the most precise results previously  obtained for this model are those presented in Ref.~\cite{Gautier}, which are also valid up to the NLO in the large N expansion, but only at the leading IR order. We have also shown that our results coincide with the ones of Ref.~\cite{Gautier}, when expanded up to the corresponding order. Beyond the leading IR order, as we have emphasized along the main text, a consistent treatment of the UV sector becomes necessary. The use of the Euclidean path integral (which is simpler than its {\it in-in} counterpart) together with the double perturbative expansion (in  $\sqrt\lambda$ and $1/N$) performed in our calculations, allowed us to further include the contribution of the UV modes and to consistently take into account the renormalization process. Moreover, in this framework, the precision of the calculation can be systematically improved  by computing higher order corrections. 

There are many  directions in which the present work can be extended. 
An interesting generalization of our calculations that we leave for future work is the study of the corrections in the case of a tree level double-well potential. As customarily done in Minkowski spacetime, it would be definitively interesting to use the Euclidean approach, together with the appropriate analytical continuation, to perform calculations in nonstationary situations  in a fixed de Sitter background spacetime. For instance, to develop a systematic way of computing corrections to the effective action or to the  expectation value of the energy-momentum tensor of the quantum scalar fields, which is important for studying the backreaction of the quantum fields on the dynamics of the spacetime geometry. Furthermore, it would be valuable to extend the Euclidean techniques to other models, such as theories with derivative interactions, with fermionic and/or gauge fields, and to study metric perturbations around a de Sitter (or quasi de Sitter) background. 
 
\acknowledgments
 The work of FDM and LGT has been supported by CONICET and ANPCyT. LGT was also partially supported by the ICTP/IAEA Sandwich Training Educational Programme and UBA. DNL was supported by ICTP during the initial stage of this work. FDM would like to thank ICTP for hospitality during part of this work. Some algebraic computations were assisted by the software Cadabra \cite{Cadabra}.
 
\appendix

\section{Effective Potential}\label{app-eff-pot}

In this Appendix we examine the quadratic part of the effective potential and relate it to the variance of the zero modes. We start by defining the effective action for this theory,
\begin{equation}
 \Gamma[\bar{\phi}_0,\hat{\bar{\phi}}] = W[J_0,\hat{J}] - \int_x \left( \bar{\phi}_{0a} J_{0a} + \hat{\bar{\phi}}_{a}(x) \hat{J}_{a}(x) \right),
 \label{EA}
\end{equation}
with $W[J_0,\hat{J}] = -\log(Z[J_0,\hat{J}])$ the generating functional of connected diagrams, and where 
\begin{eqnarray}
  \bar{\phi}_{0a} &=& \frac{\delta W[J_0,\hat{J}]}{\delta J_{0a}}, \label{barphi-dWdJ} \\
  \hat{\bar{\phi}}_a &=& \frac{\delta W[J_0,\hat{J}]}{\delta \hat{J}_a},
\end{eqnarray}
define the ``classical'' fields. The Effective Potential is obtained by evaluating the effective action at a constant field, that is $\hat{\bar{\phi}} = 0$, which in turn demands that $\hat{J} = 0$, and then dividing by the space volume $V_d$:
\begin{equation}
 V_d \, V_{eff}(\bar{\phi}_0) = \Gamma[\bar{\phi}_0,0] = W[J_0,0] - \bar{\phi}_{0a} J_{0a}. \label{Veff-1}
\end{equation}
With the purpose of calculating the quadratic term of $V_{eff}(\bar{\phi}_0)$ as a function of $\bar{\phi}_0$, we perform the following expansion,
\begin{equation}
\Gamma[\bar{\phi}_0,0] = \Gamma[0,0] + \frac{1}{2} \frac{\delta^2 \Gamma[\bar{\phi}_0,0] }{\delta \bar{\phi}_{0a} \delta \bar{\phi}_{0b}} \Bigg|_{\bar{\phi}_0 = 0} \bar{\phi}_{0a} \bar{\phi}_{0b} + \dots, \label{EA-taylor}
\end{equation}
where the linear term vanishes for $\bar{\phi}_0 = 0$, as it can be seen by differientiating Eq.~\eqref{Veff-1} with respect to $\bar{\phi}_0$,
\begin{equation}
 \frac{\delta \Gamma[\bar{\phi}_0,0] }{\delta \bar{\phi}_{0a}} = - J_{0a},
\end{equation}
and taking into account that in the symmetric phase, the mean field $\bar{\phi}_0$ vanishes if and only if $J_0=0$. Taking another derivative of the previous expression but now with respect to $J_0$, we obtain
\begin{equation}
 \delta_{ab} = - \frac{\delta^2 \Gamma[\bar{\phi}_0,0] }{\delta J_{0b} \delta \bar{\phi}_{0a}} = - \frac{\delta{\bar{\phi}_{0c}}}{\delta J_{0b}} \, \frac{\delta^2 \Gamma[\bar{\phi}_0,0] }{\delta \bar{\phi}_{0c} \delta \bar{\phi}_{0a}},
\end{equation}
where we used the chain rule for the second equality. Now, differientiating Eq.~\eqref{barphi-dWdJ} with respect to $J_0$ gives,
\begin{equation}
 \frac{\delta{\bar{\phi}_{0c}}}{\delta J_{0b}} = \frac{\delta^2 W[J_0,0]}{\delta J_{0b} \delta J_{0c}}, 
\end{equation}
which, inserted in the previous expression leads to the conclusion that
\begin{equation}
 \frac{\delta^2 \Gamma[\bar{\phi}_0,0] }{\delta \bar{\phi}_{0a} \delta \bar{\phi}_{0b}} = - \left( \frac{\delta^2 W[J_0,0]}{\delta J_{0a} \delta J_{0b}} \right)^{-1}.
\end{equation}
We now have to evaluate for $\bar{\phi}_0 = 0$ ($J_0 = 0$),
\begin{equation}
\frac{\delta^2 W[J_0,0]}{\delta J_{0a} \delta J_{0b}} \Bigg|_{J_0 = 0} = - \frac{1}{Z[0,0]} \frac{ \delta^2 Z [J_0,0] }{\delta J_{0a} \delta J_{0b}} \Bigg|_{J_0} = -
\langle \phi_{0a} \phi_{0b} \rangle,
\end{equation}
allowing us to identify the exact two-point functions 
of the zero modes $\langle \phi_{0a} \phi_{0b} \rangle$. In the symmetric phase we expect any rank-2 tensor with respect to the internal $O(N)$ indeces to be proportional to the identity $\delta_{ab}$. Therefore inverting the previous quantity is straightforward,
\begin{equation}
 \frac{\delta^2 \Gamma[\bar{\phi}_0,0] }{\delta \bar{\phi}_{0a} \delta \bar{\phi}_{0b}} = \delta_{ab} \frac{N}{\langle \phi_0^2 \rangle},
\end{equation}
where we have expressed the result in terms of the variance of $|\phi_0|$, i.e. $\langle \phi_0^2 \rangle = \delta_{ab} \langle \phi_{0a} \phi_{0b} \rangle$. Finally, we replace this last expression in Eq.~\eqref{EA-taylor} and we divide by $V_d$, in order to obtain the effective potential up to quadratic order, Eq.~\eqref{eff-pot-quad}.

\section{Functional derivatives of $\hat{Z}_{f}[\hat{J}]$}\label{app-deriv-Zf}

Using that $\hat{Z}_{f}[\hat{J}]$ is a free generating functional, its functional derivatives evaluated at $\hat{J} = 0$ can be easily expressed in terms of the free UV propagator $\hat{G}(x,x')$. The only somewhat tricky part is to keep track of the $O(N)$-indeces. The useful expressions are
\begin{equation}
\frac{ \delta^2 \hat{Z}_{f}[\hat{J}] }{\delta \hat{J}_a(x_1) \delta \hat{J}_b(x_2)} \Bigg|_{\hat{J}=0} = \hat{G}^{(m)}(x_1,x_2) \delta_{ab}, 
\label{Zf2}
\end{equation}
\begin{eqnarray}
\frac{ \delta^4 \hat{Z}_{f}[\hat{J}] }{\delta \hat{J}_a(x_1) \delta \hat{J}_b(x_2) \delta \hat{J}_c(x_3) \delta \hat{J}_d(x_4)} \Bigg|_{\hat{J}=0} &=& \hat{G}^{(m)}(x_1,x_2) \hat{G}^{(m)}(x_3,x_4) \delta_{ab} \delta_{cd} \notag \\
&&+ \hat{G}^{(m)}(x_1,x_3) \hat{G}^{(m)}(x_2,x_4) \delta_{ac} \delta_{bd} \notag \\
&&+ \hat{G}^{(m)}(x_1,x_4) \hat{G}^{(m)}(x_2,x_3) \delta_{ad} \delta_{bc}. \label{Zf4}
\end{eqnarray}
In the case of the sixth derivative, it is not necessary to write down the most general expression for six different points, since we only need particular cases with some of them evaluated in coincidence. The two particular cases we need are
\begin{eqnarray}
\Biggl[ (N+4) \delta_{cd} \delta_{ef} &+& 4\frac{A_{cdef}}{(N+2)} \Biggr] \iint_{x,x'} \frac{ \delta^6 \hat{Z}_{f}[\hat{J}] }{\delta \hat{J}^6_{abcdef}(x_1,x_2,x,x,x',x')} \Bigg|_{\hat{J}=0} \notag \\
&=& N \Biggl[ (N+2)^2 V_d^2 [\hat{G}^{(m)}]^2 + 2(N+8) \iint_{x,x'} \hat{G}^{(m)}(x,x')^2 \Biggr] \delta_{ab} \hat{G}^{(m)}(x_1,x_2) \notag \\
&&+ 4 (N+2)^2 \delta_{ab} V_d [\hat{G}^{(m)}] \int_x \hat{G}^{(m)}(x_1,x) \hat{G}^{(m)}(x,x_2) \notag \\
&&+ 8(N+8) \delta_{ab} \iint_{x,x'} \hat{G}^{(m)}(x_1,x) \hat{G}^{(m)}(x',x_2) \hat{G}^{(m)}(x,x'), \label{Zf6a}
\end{eqnarray}
and
\begin{eqnarray}
\delta_{cd} \delta_{ef} \int_{x} \frac{ \delta^6 \hat{Z}_{f}[\hat{J}] }{\delta \hat{J}^6_{abcdef}(x_1,x_2,x,x,x,x)} \Bigg|_{\hat{J}=0} &=& N (N+2) \delta_{ab} V_d [\hat{G}^{(m)}]^2 \hat{G}^{(m)}(x_1,x_2) \label{Zf6b} \\
&&+ 4(N+2) \delta_{ab} [\hat{G}^{(m)}] \int_x \hat{G}^{(m)}(x_1,x) \hat{G}^{(m)}(x,x_2). \notag
\end{eqnarray}


\section{Renormalization}\label{app-ren}

The renormalization process is performed by the addition of two counterterms in the action, $\int d^dx \sqrt{g} \delta m^2 \phi_a \phi_a/2$ and $\int d^dx \sqrt{g} \delta \lambda (\phi_a \phi_a)^2/8N$. It is safe to assume that, as in the usual perturbative case, $\delta m^2 \sim \lambda$ and $\delta \lambda \sim \lambda^2$, therefore at NNLO we need to consider terms with $\delta m^2$, $(\delta m^2)^2$ and $\delta \lambda$. This leads to the following new contributions to the generating functional,
\begin{eqnarray}
 \Delta Z[J_0,\hat{J}] &=& Z_0[J_0] \hat{Z}_{f}[\hat{J}] \Biggl\{ \frac{V_d \delta m^2}{2} \left( \langle \phi_0^2 \rangle_0 - \frac{ \delta^2 Z_0 [J_0] }{\delta J_{0a} \delta J_{0b}} \delta_{ab} \right) \\
 &&\,\,\,\,\,\,\,\,\,\,\,\,\,\,\,\,\,\,\,\,\,\,\,\,\,\,\,\,\,\,\,\,\,\,\, + \left[ \frac{V_d \delta \lambda}{8N} \langle \phi_0^4 \rangle_0 + \frac{V_d^2 (\delta m^2)^2}{4} \left( \langle \phi_0^2 \rangle_0^2 - \frac{\langle \phi_0^4 \rangle_0}{2} \right)  \right] \Biggr\} \notag \\
 &&+\hat{Z}_f[\hat{J}] \Biggl\{ - \frac{V_d^2 (\delta m^2)^2}{4} \langle \phi_0^2 \rangle_0 \frac{ \delta^2 Z_0 [J_0] }{\delta J_{0a} \delta J_{0b}} \delta_{ab} \notag \\
 &&\,\,\,\,\,\,\,\,\,\,\,\,\,\,\,\,\,\,\,\,\,\,\,\, +\left( \frac{V_d^2 (\delta m^2)^2}{8} - \frac{V_d \delta \lambda}{8N} \right) \frac{\delta^4 Z_0 [J_0] }{\delta J_{0a} \delta J_{0b} \delta J_{0c} \delta J_{0d}} \delta_{ab} \delta_{cd} \Biggr\} \notag \\
 &&+ Z_0[J_0] \frac{\delta m^2}{2} \Biggl[ \hat{Z}_f[\hat{J}] N V_d [\hat{G}^{(m)}] - \int_x \frac{ \delta^2 \hat{Z}_f[\hat{J}] }{\delta \hat{J}_a(x) \delta \hat{J}_b(x)} \delta_{ab} \Biggr] \notag \\
 &&+ \frac{\lambda}{4N} \frac{V_d \delta m^2}{2} A_{abcd} \int_x \frac{ \delta^2 \hat{Z}_f[\hat{J}] }{\delta \hat{J}_c(x) \delta \hat{J}_d(x)} \notag \\
 &&\,\,\,\,\,\,\,\,\,\,\,\,\,\,\,\,\,\,\,\,\,\,\,\,\, \times \Biggl[ \frac{\delta^4 Z_0 [J_0] }{\delta J_{0a} \delta J_{0b} \delta J_{0e} \delta J_{0f}} - \langle \phi_0^2 \rangle_0 \frac{ \delta^2 Z_0 [J_0] }{\delta J_{0a} \delta J_{0b}} \delta_{ab} \Biggr]. \notag
\end{eqnarray}
Tracking these terms into the calculation of both the UV and IR two-point functions, produce the following contributions:
\begin{eqnarray}
 \Delta\left(\langle \hat{\phi}_{a}(x) \hat{\phi}_{b}(x') \rangle\right) = \delta_{ab} \Biggl[ 1 -\frac{\lambda}{4N^2}(N+2) V_d \left( \langle \phi_0^4 \rangle_0 - \langle \phi_0^2 \rangle_0^2 \right) \Biggr] \delta m^2 \frac{\partial \hat{G}^{(m)}(x,x')}{\partial m^2},
\end{eqnarray}
to be added to Eq.~\eqref{2-pt-UV2}, and
\begin{eqnarray}
 \Delta\left(\langle \phi_{0a} \phi_{0b} \rangle\right) &=& \frac{\delta_{ab}}{N}\Biggl\{ \frac{V_d\delta m^2}{2} \left( \langle \phi_0^2 \rangle_0^2 - \langle \phi_0^4 \rangle_0 \right) \\
 &&+ \left( \langle \phi_0^6 \rangle_0 - \langle \phi_0^2 \rangle_0 \langle \phi_0^4 \rangle_0 \right) \Biggl[ \frac{V_d^2 (\delta m^2)^2}{8} - \frac{V_d \delta \lambda}{8N} + \frac{\lambda}{8N}(N+2) \delta m^2 V_d^2 [\hat{G}^{(m)}] \Biggr] \notag \\
 &&+ \left( \langle \phi_0^2 \rangle_0^3 - \langle \phi_0^2 \rangle_0 \langle \phi_0^4 \rangle_0 \right) \Biggl[ \frac{V_d^2 (\delta m^2)^2}{4} + \frac{\lambda}{4N}(N+2) \delta m^2 V_d^2 [\hat{G}^{(m)}]  \Biggr] \Biggr\}, \notag
\end{eqnarray}
to be added to a previous step of Eq.~\eqref{2-pt-IR-ren} (not shown), which is just like Eq.~\eqref{2-pt-IR-ren} with the $ren$ and $fin$ labels removed.

With these additions coming from the counterterms, it is straightforward to see that the choices Eqs.~\eqref{delta-m-euclidean} and \eqref{delta-lambda-euclidean} render the results finite, leading to the renormalized expressions Eqs.~\eqref{2-pt-UV-ren} and \eqref{2-pt-IR-ren}.

%
%
%
%
\section{Integrals of the Euclidean UV propagator}\label{app-int-UVprop}

The UV propagator can be written as
\begin{equation}
 \hat{G}^{(m)}(x,x') = H^d \sum_{\vec{L} \neq 0} \frac{Y_{\vec{L}}(x) Y^*_{\vec{L}}(x')}{H^2 L(L+d-1)+m^2}.
\end{equation}
Then, we have
\begin{eqnarray}
 \int_{x} \hat{G}^{(m)}(x,x) &=& H^d \sum_{\vec{L} \neq 0} \frac{\left( \int_x Y_{\vec{L}}(x) Y^*_{\vec{L}}(x) \right)}{H^2 L(L+d-1)+m^2} \notag \\
 &=& \sum_{\vec{L} \neq 0} \frac{1}{H^2 L(L+d-1)+m^2},
\end{eqnarray}
where we have used the orthogonality relation of the spherical harmonics
\begin{equation}
 \int_x Y_{\vec{L}}(x) Y^*_{\vec{L'}}(x) = H^{-d} \delta_{\vec{L} \vec{L'}}.
\end{equation}

Now focusing on the NNLO contributions to the UV part of the two-point function, we have
\begin{equation}
 \int_{z} \hat{G}^{(m)}(x,z) \hat{G}^{(m)}(z,x').
\end{equation}
Expanding in spherical harmonics, integrating and using the orthogonality relation we obtain
\begin{eqnarray}
H^{-2d} \sum_{\vec{L} \neq 0, \vec{L'} \neq 0}&&  \frac{Y_{\vec{L}}(x) \left( \int_z Y^*_{\vec{L}}(z) Y_{\vec{L'}}(z) \right)Y^*_{\vec{L'}}(x')}{[H^2 L(L+d-1)+m^2] [H^2 L'(L'+d-1)+m^2]} \notag \\
 &&=H^{-d} \sum_{\vec{L} \neq 0} \frac{Y_{\vec{L}}(x) Y^*_{\vec{L'}}(x')}{[H^2 L(L+d-1)+m^2]^2}.
\end{eqnarray}
Now, we notice that 
\begin{eqnarray}
\frac{1}{[H^2 L(L+d-1)+m^2]^2} = - \frac{\partial}{\partial m^2} \left[ \frac{1}{H^2 L(L+d-1)+m^2} \right],
\end{eqnarray}
and therefore, under the assumption that we can pull the derivative out of the series, we conclude that
\begin{equation}
\int_{z} \hat{G}^{(m)}(x,z) \hat{G}^{(m)}(z,x') = - \frac{\partial \hat{G}^{(m)}(x,x')}{\partial m^2}.
\end{equation}
Similarly, it can be shown that 
\begin{eqnarray}
 \iint_{y,z} \hat{G}^{(m)}(x,y) \hat{G}^{(m)}(y,z) \hat{G}^{(m)}(z,x') = \frac{1}{2} \frac{\partial^2 \hat{G}^{(m)}(x,x')}{\partial (m^2)^2}.
\end{eqnarray}

\section{UV Propagators (analytical continuation and IR behavior)} \label{app-UVprop}
 The full Euclidean propagator with mass $m^2$ in d-dimensions is
\begin{eqnarray}\label{Fulprop}
G^{(m)}(x,x')=\frac{ H^{d-2}\Gamma (\frac{d-1}{2}-\nu_d) \Gamma (\frac{d-1}{2}+\nu_d) }{(4\pi)^{d/2} \Gamma \left(\frac{d}{2}\right)}  \,_2F_1\left(\frac{d-1}{2}-\nu_d,\frac{d-1}{2}+\nu_d;\frac{d}{2};s\right),
\end{eqnarray} where $\nu_d=\sqrt{\frac{(d-1)^2}{4}-\frac{m^2}{H^2}}$,   $s=(1+z)/2$, with  $z = \delta_{AB} X^A (x)X^B(x')$  (using the embedding in $d+1$-dimensional Euclidean space and cartesian coordinates). After the analytical continuation $X^0\to iX^0$, and using comoving coordinates we can write this de Sitter invariant variable as $s=(1+z)/2=1-r/4$ where  $r=[-(\eta-\eta')^2+|\vec{x}-\vec{x'}|^2]/\eta\eta'$.
Now, we can use the $i\epsilon$ prescription to obtain the Feynman propagator $z\to z-i\epsilon$. In Lorentzian spacetime, unlike in Euclidean space, distances between points can be arbitrarily large. In the limit $r\to+\infty$, i.e. at large spatial separations or late times, the massive free propagator decays as
\begin{equation}
 G_{F}^{(m)}(r) \sim r^{-m^2/(d-1)H^2}.
\end{equation}

In order to consider the massless case, we first obtain the Euclidean UV propagator from Eq.~\eqref{Fulprop}, by subtracting the contribution from the zero modes:
\begin{equation}\label{FulpropUV}
\hat{G}^{(m)}={G}^{(m)}-G_0^{(m)}
\end{equation} with $G_0^{(m)}=1/(V_d m^2)$.
For $m\to 0$, we get
\begin{eqnarray}
\hat{G}^{(0)}(r)&=&\frac{H^{d-2}}{(4\pi)^{d/2}}\left\{\frac{ s \Gamma (d) \,
   _3F_2\left(1,1,d;2,\frac{d}{2}+1;1-r/4\right)}{\Gamma
   \left(\frac{d+2}{2}\right)}-\frac{ \left(h_{d-2} \Gamma (d)+\Gamma
   (d-1)\right)}{(d-1) \Gamma \left(\frac{d}{2}\right)}\right\}, \,\,\,\,\,\,\,\,
\end{eqnarray} where $h_n=1+1/2+\dots +1/n$ is the Harmonic number. For $d=4$ it reduces to 
 \begin{equation}
\hat{G}^{(0)}(r)=H^2\left\{\frac{14 (1-r/4)-(3r/2) \log (r/4)-11}{12 \pi ^2 r}\right\}.
 \end{equation}
 Then, using the $i\epsilon-$prescription,  the corresponding massless Feynman propagator is
 \begin{equation}
\hat{G}^{(0)}_F(r) =\frac{H^2}{(4\pi)^2} \left\{\frac{4}{r+i\epsilon} -2\log \left(\frac{r+i\epsilon}{4}\right)-\frac{14}{3}\right\},
 \end{equation} 
 showing that when $r \to +\infty$
\begin{equation}
 \hat{G}_{F}^{(0)}(r) \sim \log(r).
\end{equation}
Analogously, it can be shown that the derivatives with respect to the mass pick up further powers of $\log(r)$:
\begin{equation}
 \frac{\partial^k \hat{G}_{F}^{(m)}(r)}{\partial (m^2)^k} \Bigg|_{0} \sim \log(r)^{k+1}.
\end{equation}

\section{Double expansion and diagrammatics (the massless case)}\label{app-Double-expansion}

In this Appendix we  focus on the corrections to the UV propagator and the associated diagrammatics   for the case of fields with vanishing tree-level masses.
The zero modes are treated nonperturbatively  using the exact  Euclidean  path integrals  in the absence of the nonzero modes.  
On  the one hand, from these  path integrals  we have learnt that each power of $\phi_{0,a}$  scales  as $\lambda^{-1/4}$, while each $\hat{\phi}_{a}$   does not add any factor of  $\lambda$ (i.e., it counts as $\lambda^0$). On the other hand, we know that   in  the massless limit   ($m^2\to 0$) the  free propagators  in the Lorentzian spacetime  increase at large distances, and that  a derivative with respect to  $m^2$ of the propagators  adds a  logarithmic-growth factor (see Appendix \ref{app-UVprop}).  

For assessing the importance of each diagram after the analytical continuation to the Lorentzian spacetime  we need to understand how the behavior of each of them at large distances would be. For this, it is very useful to note  that equations like \eqref{G2-dGdm} and \eqref{G3-d2Gdm2} hold  in general, namely
\begin{eqnarray}
\int ...\int_{x_2,..,x_{k-1}} \hat{G}^{(m)}(x_1,x_2)... \hat{G}^{(m)}(x_{k-1},x_k) \propto   \frac{\partial^{k-2}\hat{G}^{(m)}(x,x')}{\partial (m^2)^{k-2}}.\label{Gk}
\end{eqnarray} 
Therefore,   by representing each logarithmic-growth factor   by  $y$ (i.e., using the notation of Appendix \ref{app-UVprop}, $y\sim \log r$) we see   the right hand side scales as $y^{k-1}$.

\begin{figure}[htbp]
\begin{center}
\begin{eqnarray*} 
\parbox{30mm}{\begin{fmfgraph}(80,30)\fmfkeep{resum}
\fmfleft{i} \fmfright{o} \fmftop{t} \fmf{plain}{i,v1,o} \fmf{dashes,left,tension=0}{v1,t,v1}  
\end{fmfgraph}} 
\end{eqnarray*} 
\caption{Feynman diagram contributing to the UV propagator. Free UV propagators are represented by a solid line, while the full variance of the zero modes in the absence of the other modes is represented by a dashed line.}
\label{Fig3}
\end{center}
\end{figure}
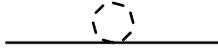

\begin{figure}[htbp]
\begin{center}
\begin{eqnarray*}
&&\parbox{45mm}{\begin{fmfgraph}(120,30)\fmfkeep{2loop-resum}
\fmfstraight
\fmfleft{i} \fmfright{o} \fmftop{ti,t1,t2,to} \fmf{plain}{i,v1,v2,o} \fmf{dashes,left,tension=0}{v1,t1,v1} \fmf{dashes,left,tension=0}{v2,t2,v2} 
\end{fmfgraph}} + \parbox{58mm}{\begin{fmfgraph}(160,30)\fmfkeep{3loop-resum}
\fmfstraight
\fmfleft{i} \fmfright{o} \fmftop{ti,t1,t2,t3,to} \fmf{plain}{i,v1,v2,v3,o} \fmf{dashes,left,tension=0}{v1,t1,v1} \fmf{dashes,left,tension=0}{v2,t2,v2} \fmf{dashes,left,tension=0}{v3,t3,v3} 
\end{fmfgraph}} + \dots
\end{eqnarray*}
\caption{Feynman diagrams contributing to the UV propagator, which become of the same order as the one in Fig.~\ref{Fig3} when $\sqrt{\lambda}y\sim1$. The first one scales as $y(\sqrt{\lambda}y)^2$, the second one as $y(\sqrt{\lambda}y)^3$ and so on.}
\label{Fig4}
\end{center}
\end{figure}
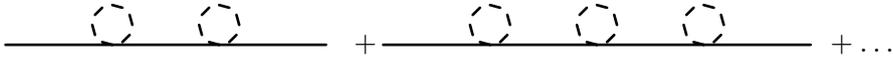

In order to draw   Feynman diagrams, we use  a dash line for  the variance of the zero modes (which  is always computed nonperturbatively using the  exact  path integral for the zero modes in isolation and scales  as $1/\sqrt{\lambda}$) and a solid line  for the  free  UV propagator (which depending on the diagram may contribute with a factor of  $y$).
Hence, using that for each vertex  we have  a factor of $\lambda/N$,  one can conclude that    the  diagram  in Fig.~\ref{Fig3}  scales as $y(\sqrt{\lambda}y)$  and  becomes of the same order as the free UV propagator (which scales as $y$) when  $\sqrt{\lambda}y\sim 1$,   indicating a break down of the perturbation theory  when the distance is sufficiently large.
  Actually, there are more diagrams that also become of the same order in that case, as, for instance, those shown in Fig.~\ref{Fig4}. Indeed, it is not difficult to see that the one with $n$-vertices scales as $y(\sqrt{\lambda}y)^n$, representing a relative correction that goes as $(\sqrt{\lambda}y)^n$.  We were able to perform the resummation of these diagrams in section \ref{Ressum}. 

\begin{figure}[htbp]
\begin{center}
\begin{eqnarray*}
\parbox{40mm}{\begin{fmfgraph}(80,50)\fmfkeep{det}
\fmfleft{i} \fmfright{o} \fmftop{t} \fmf{plain}{i,v1,o} \fmf{dashes,left,tension=0.001}{v1,v2,v1} \fmf{plain,left,tension=0.001}{v2,t,v2}  
\end{fmfgraph}}
\parbox{40mm}{\begin{fmfgraph}(80,50)\fmfkeep{1loop-hatphi4}
\fmfleft{i} \fmfright{o} \fmftop{t} \fmf{plain}{i,v1,o} \fmf{plain,left,tension=0}{v1,t,v1}  
\end{fmfgraph}} 
\end{eqnarray*}
\caption{Feynman diagrams contributing to the UV propagator that are to be  included perturbatively. Both diagrams scale as $(\sqrt{\lambda}y)^2$.}
\label{Fig5}
\end{center}
\end{figure}
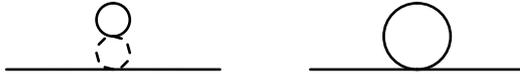

Now, in order to see that the rest of the contributions can be treated perturbatively, let us analyze each of them separately, according to the type of vertex. Let us start with other corrections associated to the bi-quadratic interaction term of Eq.~\eqref{biq}.  The leading order  diagram involving  this vertex which is  not included in the resummation is the one on the left  in Fig.~\ref{Fig5}, and gives a relative correction that scale  as  $\sqrt{\lambda}(\sqrt{\lambda}y)$. Then, for $\lambda$ sufficiently small it can be considered as a small correction to the ones of Fig.~\ref{Fig3} at all times. The other type of correction that also contributes  up to the NLO in $1/N$ involves   the last  interaction term  in Eq.~\eqref{S-int-tilde} and the corresponding diagram is the one   in Fig.~\ref{Fig5}  to the right and   gives a correction relative to the free propagator that  also goes as $\sqrt{\lambda} (\sqrt{\lambda}y)$.


\begin{fmffile}{diagram3}
\fmfset{dot_len}{0.7mm}

\end{fmffile}

\begin{fmffile}{diagram4}
\end{fmffile}

\begin{fmffile}{diagram5}

\end{fmffile}

\section{Derivative of $[\det \hat{G}_{rs}^{(m)}]^{1/2}$}\label{app-deriv-det}

From the well known identity
\begin{equation}
 \log \det (\hat{G}_{rs}^{(m)}) = \textup{Tr} \log (\hat{G}_{rs}^{(m)}),
\end{equation}
we have
\begin{equation}
 \log [\det \hat{G}_{rs}^{(m)}]^{1/2} = - \frac{1}{2} \delta_{ab} \int_x \log (\hat{G}_{ab}^{-1}(x,x)).
\end{equation}
Diferentiating at both sides with respect to $m_{ij}^2$ we then obtain
\begin{eqnarray}
 &&\frac{1}{[\det \hat{G}_{rs}^{(m)}]^{1/2}} \frac{\partial [\det \hat{G}_{rs}^{(m)}]^{1/2}}{\partial m_{ij}^2} \notag \\
 &&\,\,\,\,\,\,\,\,\,\,\,\,= - \frac{1}{2} \delta_{ab} \iint_{x,x'} \hat{G}_{ac}(x,x') \frac{\partial \hat{G}^{-1}_{cb}(x',x)}{\partial m_{ij}^2},
\end{eqnarray}
while from Eq.~\eqref{inv-hatG} we know that
\begin{equation}
\frac{\partial \hat{G}^{-1}_{ab}(x,x')}{\partial m_{ij}^2} = \delta_{ai} \delta_{bj} \delta^{(d)}(x-x'), 
\end{equation}
and therefore
\begin{eqnarray}
 &&\frac{1}{[\det \hat{G}_{rs}^{(m)}]^{1/2}} \frac{\partial [\det \hat{G}_{rs}^{(m)}]^{1/2}}{\partial m_{ij}^2} = - \frac{1}{2} \delta_{ij} V_d [\hat{G}],
\end{eqnarray}
where we used that $[\hat{G}_{ab}] = \delta_{ab} [\hat{G}]$ in the symmetric phase. Tracing the remaining free indeces $ij$ proves 
Eq.~\eqref{partial-det}.

\section{Expanding the UV resummed result in $\sqrt{\lambda}$}\label{app-resum-exp}

As a cross-check, we expand back the UV resummed two-point functions 
of the UV modes, Eq.~\eqref{UV-prop-NLO-N-1}, and compare it to the perturbative one, Eq.~\eqref{2-pt-UV-ren}, for $m=0$. All the UV propagators with dynamical masses must be expanded in $\sqrt{\lambda}$, up to the needed order. For example
\begin{eqnarray}
 \hat{G}^{(m_{dyn,0})}(x,x') &=& \hat{G}^{(0)}(x,x') + \sqrt{\frac{\lambda}{2V_d}} \frac{\partial \hat{G}^{(0)}(x,x')}{\partial m^2} + \frac{1}{2} \frac{\lambda}{2V_d} \frac{\partial^2 \hat{G}^{(0)}(x,x')}{\partial (m^2)^2} + \mathcal{O}(\lambda^{3/2}), \,\,\,\,\,\,\,\,
\end{eqnarray}
and 
\begin{eqnarray}
 2 \hat{G}^{(\sqrt{3} m_{dyn,0})}(x,x') - 2 \hat{G}^{(m_{dyn,0})}(x,x') &=& 4 \sqrt{\frac{\lambda}{2V_d}} \frac{\partial \hat{G}^{(0)}(x,x')}{\partial m^2} \\
 &&+ 8 \frac{\lambda}{2V_d} \frac{\partial^2 \hat{G}^{(0)}(x,x')}{\partial (m^2)^2} + \mathcal{O}(\lambda^{3/2}), \notag
\end{eqnarray}
and so on. The resulting expression up to order $\lambda$ and $1/N$ is precisely Eq.~\eqref{2-pt-UV-massless-NLO-N-1-expanded}.

\end{document}